\documentclass [11pt]{article}

\usepackage{jheppub}

\usepackage[utf8]{inputenc}

\usepackage{amsfonts}

\usepackage{amsmath,amssymb}

\usepackage{mathtools}
\usepackage{subcaption}
\usepackage{graphicx}
\graphicspath{ {./figures/}}

\usepackage[space]{grffile}
\usepackage{mathtools}

\usepackage{simplewick}

\usepackage{array}

\usepackage{float}

\usepackage{color}

\usepackage{mathrsfs}

\usepackage{ytableau}

\usepackage{tikz}
\usepackage{tikz-feynman}
\usepackage{contour}
\usepackage{slashed}
\usepackage{booktabs}

\newcommand{\cO}{\mathcal{O}}
\newcommand{\cT}{\mathcal{T}}
\newcommand{\cN}{\mathcal{N}}

\newcommand{\be}{\begin{eqnarray}\displaystyle}
\newcommand{\ee}{\end{eqnarray}}

\newcommand{\myRho}{{\mathfrak{r}}}

 \definecolor{verde}{rgb}{0,0.7,0.2}

\renewcommand{\Re}{\operatorname{Re}}
\renewcommand{\Im}{\operatorname{Im}}

\newcommand{\pder}[1]{\partial_#1}

\title{\boldmath Bounds on scattering of neutral Goldstones}

\author[a]{Francesca Acanfora,}
\author[d,e]{Andrea Guerrieri,}
\author[b,c]{Kelian H\"aring,}
\author[f]{and Denis Karateev}

\affiliation[a]{Galileo Galilei Institute for Theoretical Physics, Largo Enrico Fermi 2, I-50125 Firenze, Italy}

\affiliation[b]{Theoretical Physics Department,
	CERN, 1211 Geneva 23, Switzerland}

\affiliation[c]{Fields and Strings Laboratory, Institute of Physics École Polytechnique Fédéral de Lausanne (EPFL)
	\\ Route de la Sorge, CH-1015 Lausanne, Switzerland}

\affiliation[d]{Perimeter Institute for Theoretical Physics, 31 Caroline St N Waterloo, Ontario N2L 2Y5, Canada}

\affiliation[e]{Dipartimento di Fisica e Astronomia, Universita degli Studi di Padova \\
\& Istituto Nazionale di Fisica Nucleare, Sezione di Padova, via Marzolo 8, 35131 Padova, Italy}

\affiliation[f]{
	D\'epartment de Physique Th\'eorique, Universit\'e de Gen\`eve,\\
	24 quai Ernest-Ansermet, 1211 Gen\`eve 4, Switzerland}

\abstract{
We study the space of $2\to 2$ scattering amplitudes of neutral Goldstone bosons in four space-time dimensions. 
We establish universal bounds on the first two non-universal Wilson coefficients of the low energy Effective Field Theory (EFT) for such particles.
We reconstruct the analytic, crossing-symmetric, and unitary amplitudes saturating our bounds, and we study their physical content.
We uncover non-perturbative Regge trajectories by continuing our numerical amplitudes to complex spins.
We then explore the consequence of additional constraints arising when we impose the knowledge about the EFT up to the cut-off scale.
In the process, we improve on some aspects of the numerical $S$-matrix bootstrap technology for massless particles. 
}

\begin{document}
	\begin{flushleft}
	\hfill \parbox[c]{40mm}{CERN-TH-2023-179}
\end{flushleft}
\maketitle

\newpage
\section{Introduction}\label{introduction}

In the past decade there has been a surge of interest in studying general properties of scattering amplitudes stemming from the fundamental principles of causality, crossing, and unitarity.  One powerful method for doing this is the numerical $S$-matrix bootstrap~\cite{Paulos:2016fap, Paulos:2016but, Paulos:2017fhb}. It allows to determine model-independent bounds on physical observables and extract scattering amplitudes from them.

The case of massive particles has been widely studied in various space-time dimensions~\cite{Doroud:2018szp, He:2018uxa, Cordova:2018uop, Guerrieri:2018uew, Paulos:2018fym, Homrich:2019cbt, Cordova:2019lot, Bercini:2019vme, Gabai:2019ryw, Bose:2020shm, Bose:2020cod, Kruczenski:2020ujw,Karateev:2019ymz,Karateev:2020axc,Guerrieri:2020kcs,He:2021eqn,Guerrieri:2021tak,Chen:2021pgx,Cordova:2022pbl,Karateev:2022jdb,Chen:2022nym,EliasMiro:2022xaa,Sinha:2022crx,Antunes:2023irg,Buric:2023ykg,Marucha:2023vrn,Correia:2022dyp,Levine:2023ywq,Ghosh:2023rpj,He:2023lyy,Dersy:2023job}. Particles with spins have been studied in \cite{Hebbar:2020ukp}. Alternative approaches have been developed in~\cite{Tourkine:2021fqh, Tourkine:2023xtu,Henning:2022xlj,Fitzpatrick:2023mbt}, and progress has been made in establishing new rigorous results in~\cite{Correia:2021etg,Correia:2022dcu}.
The case of massless particles has been investigated to a lesser extent, and the numerical algorithms are less effective. Nevertheless, some intriguing results have been obtained: bounds on the quark antiquark potential~\cite{EliasMiro:2019kyf,EliasMiro:2021nul}, universal constraints on pion low energy constants~\cite{Guerrieri:2020bto}, no-go theorems for quantum gravity with supersymmetry ~\cite{Guerrieri:2021ivu,Guerrieri:2022sod}, and universal bounds on photon scattering~\cite{Haring:2022sdp}. 

In this work, we apply the numerical bootstrap to a simple physical process: the $2\to 2$ scattering amplitude of massless identical scalars in four space-time dimensions.\footnote{Stronger bounds can be obtained by including multi-particle constraints. This is an extremely hard problem in higher dimension. In 2d some progress has been made in~\cite{to_appear_multiparticles}. }
We focus on the class of IR safe amplitudes that can be described at low energies by a derivatively coupled Effective Field Theory (EFT)\footnote{Here we have only included the kinetic term and terms which contain four fields $\phi$ and up to 8 derivatives. }
\begin{multline}
	\label{eq:EFT_lagrangian}
	{\cal L}_\text{EFT} = -\dfrac{1}{2} (\partial \phi)^2+ c_4 (\partial \phi)^4 +
	c_6 (\pder{\mu} \pder{\nu} \pder{\rho}\phi) (\pder{\mu} \phi)(\pder{\nu} \phi)(\pder{\rho}\phi) +
	c_8 (\pder{\mu} \pder{\nu} \phi)^4 + \ldots
\end{multline} 
Examples of such theories include neutral Goldstone bosons of the spontaneously broken $U(1)$ symmetry, spontaneous conformal symmetry breaking, or 
co-dimension one defects.
The real coefficients $c_n$ in the potential are called the Wilson coefficients. Their mass dimension reads as $[c_n]=-n$. 

We can use the EFT Lagrangian density \eqref{eq:EFT_lagrangian} to systematically compute the $2\to 2$ scattering amplitude of Goldstones at each order in the small $s$ expansion $\mathcal{T}^{(k)}_\text{EFT}\sim \mathcal{O}(s^k)$, where $s$ is the Mandelstam variable describing the squared total energy of the process.
The EFT expansion provides the most generic parametrization of the amplitude compatible with crossing symmetry, and unitarity.
Since the Mandelstam variable $s$ is dimensionful, the validity of the EFT expansion as an \emph{approximation} of the amplitude itself depends on a scale, a priori unknown, called the EFT \emph{cutoff}.
For the moment, we will be agnostic about the cutoff. Instead, without loss of generality, we 
identify the EFT expansion with the small $s$ asymptotic expansion of the nonperturbative scattering amplitude of Goldstones\footnote{To be more precise, the small $s$ expansion is valid at fixed scattering angle $\theta$. Since $t=-\tfrac{s}{2}(1-\cos\theta)$, and $u=-\tfrac{s}{2}(1+\cos\theta)$, then $t,u$ are small. For simplicity of narration we also use the identification $O(s^n)\sim O\left(s^n \log^m (-s)\right)$, where $m\geq 1$.}
\begin{equation}
	\label{eq:soft_1}
	\mathcal{T}(s,t,u) = \sum_{k=2}^N
	\mathcal{T}^{(k)}_\text{EFT}(s,t,u) + O(s^{N+1}),
\end{equation}
where the first terms are given by
\begin{equation}
\label{eq:soft_2_3}
	\mathcal{T}^{(2)}_\text{EFT}(s,t,u) = g_2 \, (s^2+t^2+u^2),\qquad
	\mathcal{T}^{(3)}_\text{EFT}(s,t,u) = g_3\, stu,
\end{equation}
together with\footnote{The real coefficients $g_n$ introduced in the EFT expansion are simply related to the Wilson coefficients as $g_2 = 2c_4$, and $g_3 = 3c_6$. The relation between $g_4$ and $c_4$ is scheme-dependent. We work in the scheme in which the physical imaginary part is given by \eqref{eq:L_expression}, this leads to the relation $g_4 =c_8/4$. } 
\begin{multline}
	\label{eq:L_expression}
	\mathcal{T}^{(4)}_\text{EFT}(s,t,u) = g_4 \left(s^2+t^2+u^2\right)^2  - \frac{g_2^2}{480\pi^2} \Big(
	s^2 (41 s^2+t^2+u^2) \log\left(-s\sqrt{g_2} \right) +\\
	t^2 (s^2+41 t^2+u^2) \log\left(-t\sqrt{g_2} \right) +
	u^2 (s^2+t^2+41 u^2) \log\left(-u\sqrt{g_2} \right) \Big).
\end{multline}
The $s$, $t$ and $u$ are the usual Mandelstam variables which obey the relation $s+t+u=0$.
The above expressions can be easily derived without any reference to the Lagrangian by using only crossing, and unitarity as shown in appendix \ref{app:one-loop_unitarity}, where we provide the full expansion up to $\mathcal{O}(s^6)$ order. 

Let us stress that the logs in \eqref{eq:L_expression} contain the scale $\sqrt{g_2}$ which defines our coupling $g_4$. Alternatively we could define the coupling $g_4(\mu)$ with an arbitrary scale $\mu$ in the logs, namely we could use $\log(-s/\mu^2)$ instead of $\log(-s\sqrt{g_2})$ in \eqref{eq:L_expression}. Then the two observables are simply related as $g_4(\mu)= g_4-\tfrac{7g_2^2}{160 \pi^2}\log(\mu^2 \sqrt{g_2})$. In perturbation theory this is simply a change in the renormalization scheme.

When working with the scattering of massive particles there is canonical choice of the physical units: setting the mass gap $m=1$. In the case of massless particles there is no obvious choice.
In the EFT framework the natural candidate is the cutoff scale. However, 
from a non-perturbative perspective, the cutoff is not a well defined concept.
It is more natural to use the first dimensionful parameter $g_2$ in~\eqref{eq:soft_1} to set the units.
Notice that $g_2$ is a nice candidate because it is universal and notoriously positive~\cite{Adams:2006sv}.
Two dimensionless couplings are then formed as
\begin{equation}
	\label{eq:observables}
	\bar g_3 \equiv \frac{g_3}{g_2^{3/2}},\qquad
	\bar g_4 \equiv \frac{g_4}{g_2^2}.
\end{equation}
For later purpose, we will always define dimensionless `bar` variables in units of $g_2$, for example the dimensionless Mandelstam variable reads
\begin{equation}
	\bar s \equiv s\sqrt{g}_2.
\end{equation}

The first goal of our paper is
\begin{itemize}
	\item To estimate \emph{universal bounds} on $\{\bar g_3,\bar g_4\}$, and study the extremal amplitudes saturating those bounds.
	
	\item To improve the numerical non-perturbative methods for massless particles. In particular to solve the issue with slow numerical convergence observed in \cite{Haring:2022sdp}.

\end{itemize}

What can we say a priori about the allowed values of $\{\bar g_3, \bar g_4\}$? 
The first statement is that both couplings can take arbitrarily large values. 
To understand this, let us consider a situation in which all coefficients in the low energy expansion of the amplitude~\eqref{eq:soft_1} are given by a simple formula of the form
\begin{equation}
	\label{eq:coefficients_gn_weakQFT}
	g_n = k_n M^{-2n}\alpha,
\end{equation}
where $k_n$ are some order one real numbers, $\alpha$ is a dimensionless small parameter, and $M$ an arbitrary scale.
We can think of $M$ as the lightest particle in a weakly coupled QFT which we integrated out, and $\alpha$ as the dimensionless coupling of this putative UV theory. 
We provide some explicit examples in appendix \ref{app:absence_bounds}. 
Plugging \eqref{eq:coefficients_gn_weakQFT} into \eqref{eq:observables} we get
\begin{equation}
	\label{eq:observables_WC}
	\bar g_3 = \frac{k_3}{k_2^{3/2}}\,\frac{1}{\sqrt{\alpha}},\qquad
	\bar g_4 = \frac{k_4}{k_2^{2}}\,\frac{1}{\alpha}.
\end{equation}
In the weak coupling limit $\alpha \rightarrow 0$ the observables \eqref{eq:observables_WC} become infinitely large, and the logs in \eqref{eq:L_expression} become negligible.
In this regime, we conjecture that it is possible to describe the bootstrap results in perturbation theory using tree-level models.\footnote{For interesting results on weakly coupled models using ``positivity'' see for example \cite{Alberte:2021dnj,Chowdhury:2022obp,Li:2022aby,Zeng:2023jek,Hong:2023zgm,Bellazzini:2023nqj,Herrero-Valea:2022lfd,Herrero-Valea:2020wxz,Albert:2022oes,Albert:2023jtd,Ma:2023vgc,CarrilloGonzalez:2023cbf,Fernandez:2022kzi,Arkani-Hamed:2020blm,Caron-Huot:2020cmc,Tolley:2020gtv,Henriksson:2021ymi,Henriksson:2022oeu,Bellazzini:2021oaj,Bellazzini:2020cot,Sinha:2020win,Chowdhury:2021ynh,Caron-Huot:2021rmr}.} Our conjecture makes a number of observable predictions that we verify using our numerics in section \ref{sec:aymtptotic_bound}.

The complementary scenario is when both $\bar g_3,\bar g_4 \sim 1$. This is the regime in which the non-perturbative bootstrap is most powerful.
In this case the logarithms in~\eqref{eq:L_expression} are not parametrically suppressed, and the bounds are non-trivial.\footnote{In appendix~\ref{app:sum-rules} we derive the sum-rules for $\bar g_3, \bar g_4$ taking into account the logs. There it is clear that no simple bounds can be obtained with simple analytic arguments. }
In this regime the $S$-matrix bootstrap is superior to other methods like ``positivity''  because it  bounds the real part of the scattering amplitude by employing non-linear unitarity constraints, see appendix \ref{app:NL_unitarity} for details.
For the explanation of this fact using a simple analytic model see \cite{EliasMiro:2021nul}. For the comparison between the bounds obtained using the $S$-matrix bootstrap and positivity in the case of massive particles see \cite{Chen:2022nym,EliasMiro:2022xaa}.

When talking about EFT approximation of a physical phenomenon, there is an important requirement to satisfy: the \emph{separation of scales} between the IR physics and its UV completion. 
This separation is parametrized by the introduction of an additional dimensionful parameter, the cutoff of the EFT, that we denote by $M$. 
Recall, that $g_2$ was chosen to set the units. With the cutoff in mind, we can define a new dimensionless positive coupling $\xi$
\begin{equation}
	\label{eq:g2_M_relation}
	g_2 = \xi M^{-4}.
\end{equation}
In the example of the weakly coupled UV completion discussed above, we can compare this expression with \eqref{eq:coefficients_gn_weakQFT}. From this it follows that $\xi = k_2 \alpha$ is infinitesimally small for $\alpha\rightarrow 0$. Contrarily, for strongly coupled UV completions we expect $\xi\sim 1$. 
The effect of an additional gap due to the EFT cutoff on the bootstrap bounds in the case of massive particles has been discussed in \cite{EliasMiro:2022xaa}.
In~\cite{EliasMiro:2022xaa}, the authors proposed a simple algorithm to mimic the presence of the EFT cutoff in a non-perturbative amplitude by bounding its imaginary part at low energies $s\ll M^2$, and checked the consistency of the method with the EFT expectations. 
Here, we employ a refined variation of that idea using the strategy inspired by \cite{Chen:2021pgx} and we apply it to the case of massless scalars.

Imagine that an experimentalist tells us that the scattering amplitude for a massless process is well approximated by some experimental data for all $s,-t\leq M$, and that the error of this approximation is given by a function $\text{err}(s,t,u)$.
This can be implemented by the following condition
\begin{equation}
	\label{eq:condition_experimental}
	s\in[0,M^2]:\qquad
	\left|  \mathcal{T}(s,\cos\theta) - \mathcal{T}_\text{experimental}(s,\cos\theta) \right| \leq \text{err}(s,\cos\theta).
\end{equation}
Problems of this type were originally studied in \cite{Chen:2021pgx} in two dimensions in the case of form factors where instead of the experimental data the authors used the numerical data obtained using the Hamiltonian Truncation method \cite{Chen:2021bmm}.\footnote{Similar idea was recently explored in the context of pion physics in \cite{He:2023lyy}.} In this paper we will explain how to implement \eqref{eq:condition_experimental} in four dimensions. Since we do not have any experimental or numerical data for the neutral Goldstone scattering we will use as a proxi of this data the EFT representation \eqref{eq:soft_1} up to $O(s^4)$ order assuming is valid in the extended region $s\in[0,M^2]$. The error function will be estimated as our ignorance of the $O(s^5)$ order terms in this expansion. We refer to this type of study as \emph{model-dependent}. Summarizing, we define the second goal of our paper as
\begin{itemize}
	\item Determine \emph{model-dependent bounds} on $\{\bar g_3,\bar g_4\}$ by requiring that
		\begin{equation}
		\label{eq:condition_2}
		s\in[0,M^2]:\qquad
		\left|  \mathcal{T}(s,\cos\theta) - \sum_{n=2}^4 \mathcal{T}^{(n)}_\text{EFT}(s,\cos\theta) \right| \leq \text{err}(s,\cos\theta).
	\end{equation}	
\end{itemize}

The paper is organized as follows. In section \ref{sec:results} we present our numerical results on \emph{universal bounds} and discuss their physical meaning in great detail.  In section \ref{sec:machinery_scenario_1} we briefly summarize the technology of \cite{Paulos:2017fhb} and introduce our improvements for efficiently studying massless particles.  In section \ref{sec:modeldepBound} we introduce the machinery needed for obtaining {\it model-dependent} bounds defined by \eqref{eq:condition_2}.
We conclude in section \ref{sec:conclusions}. Some additional results are provided in appendices. We refer to them throughout the text in places where they become relevant.

The numerical data and a notebook to extract the amplitude and perform the fit analysis can be downloaded from the repository  \href{https://doi.org/10.5281/zenodo.8422615}{https://doi.org/10.5281/zenodo.8422615}.

\section{Universal bounds on Wilson coefficients}
\label{sec:results}

We investigate numerically the space of UV complete amplitudes of Goldstones in four dimensions. 
To this end, we use the numerical method introduced in \cite{Paulos:2017fhb} and \cite{Guerrieri:2020bto}. 
We briefly review this method in section \ref{sec:machinery_scenario_1}. There, we also introduce several crucial improvements of this method needed for studying massless particles efficiently. Below we summarize our physical results.\footnote{In this section, we present our results using the ansatz defined in \eqref{eq:ansatz_2} with its improvement. If not explicitly specified in the figures, we use $N_{\text{max}}=26$ and $n_{\text{max}}=8$ and impose unitarity constraints up to spin 200.  We refer the reader to section \ref{sec:machinery_scenario_1} for precise definition of these variables. For now, what is necessary to know is that $N_{\text{max}}$ measures the size/freedom of the ansatz.}

\begin{figure}[t]
	\centering
	\begin{subfigure}{0.49\textwidth}
		\includegraphics[width=1\linewidth]{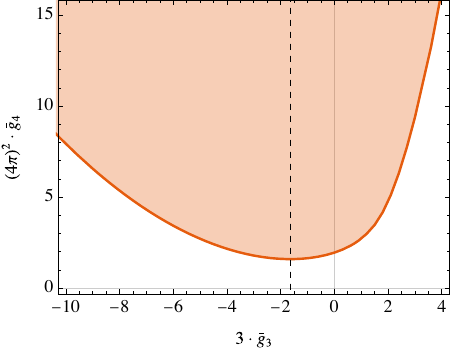}
		\caption{The vertical dashed line indicates the position of the absolute minimum of the bound. \\}
		\label{fig:bound_normal}
	\end{subfigure}
	\begin{subfigure}{0.49\textwidth}
		\includegraphics[width=1\linewidth]{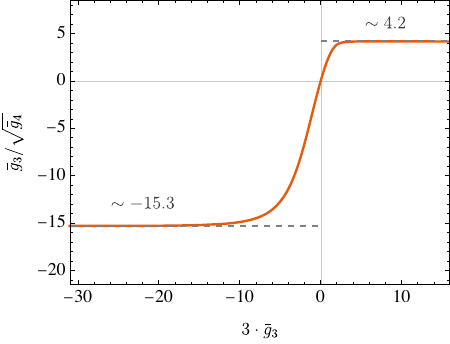}
		\caption{We plot the ratio:  $\bar{g}_3/\sqrt{\bar g_4}$ vs $\bar g_3$. Dashed lines indicate the approximate asymptotes of the bound. }
		\label{fig:bound_equivalent}
	\end{subfigure}
	\caption{The allowed region for the parameters $\bar g_3$ and $\bar g_4$ defined in \eqref{eq:observables} lies above the orange line. }
		\label{fig:bound}
\end{figure}

By using this method we determine the lower bound on the coupling $\bar g_4$ as a function of $\bar g_3$. This bound is given by the orange line in figure \ref{fig:bound_normal}. All values of the couplings $\bar g_3$ and $\bar g_4$ above the orange line are allowed. 
There is a nontrivial absolute minimum for $\bar g_4$ 
\begin{equation}
\min \bar g_4=\frac{1.58\pm 0.02}{(4\pi)^2}
\end{equation}
attained for $\bar g_3\simeq -0.54$. 
The value of $\bar g_4$ is scheme-dependent as it appears at the same order as the logs in the amplitude. We define it in terms of the physical amplitude as in equations \eqref{eq:soft_1} - \eqref{eq:L_expression}.

Looking at figure~\ref{fig:bound_normal} we observe that $\bar g_4$ is unbounded from above, while $\bar g_3$ is unbounded in both directions. 
We have already argued in section \ref{introduction} how arbitrarily large values of the Wilson coefficients can be realized by a simple mechanism. The absence of these bounds can be proven theoretically by construction (in any number of space-time dimensions) as explained in appendix \ref{app:absence_bounds}.
In figure \ref{fig:bound_equivalent} we present our bound for the ratio $\bar g_3/\sqrt{\bar g_4}$.\footnote{It is interesting to note that for the weakly coupled UV completion discussed around \eqref{eq:coefficients_gn_weakQFT} we have $\frac{\bar g_3}{\sqrt{\bar g_4}} = \frac{k_3}{\sqrt{k_2 k_4}} $ which is independent of the small coupling $\alpha$.} From figure \ref{fig:bound_equivalent}, we observe that the ratio $\bar g_3/\sqrt{\bar g_4}$ goes to a constant as $|\bar g_3|\to \infty$, and we can estimate the following bound on this ratio
\begin{equation}\label{eq:bound_ratio}
	-15.3 \lessapprox \frac{\bar g_3}{\sqrt{\bar g_4}}\lessapprox 4.2\,.
\end{equation}

When $\bar g_3=0$, our bound applies to the special case of the scattering of dilatons in $\mathcal{N}=4$ SYM on the Coulomb branch \cite{Bianchi:2016viy}. By giving a vev to one of the six scalars, we induce the spontaneous symmetry breaking of both conformal and $SU(4)$ $R$-symmetry. Fluctuations around the vev correspond to the dilaton, the remaining five real scalars correspond to the Goldstones of the unbroken $Sp(4)$ symmetry. At low energies the amplitude with only dilatons matches the expansion in \eqref{eq:soft_1} up to order $\mathcal{O}(s^4)$ included, and the mixing with the other Goldstones happens at higher orders.
The condition $\bar g_3=0$ is consequence of SUSY and the soft theorems. By looking at our bound at $\bar g_3=0$ point we report the following result
\begin{equation}
\min_{\bar g_3=0} \bar g_4 \simeq \frac{1.95}{(4\pi)^2}.
\end{equation}

\begin{figure}[t]
	\centering
	\includegraphics[scale=0.4]{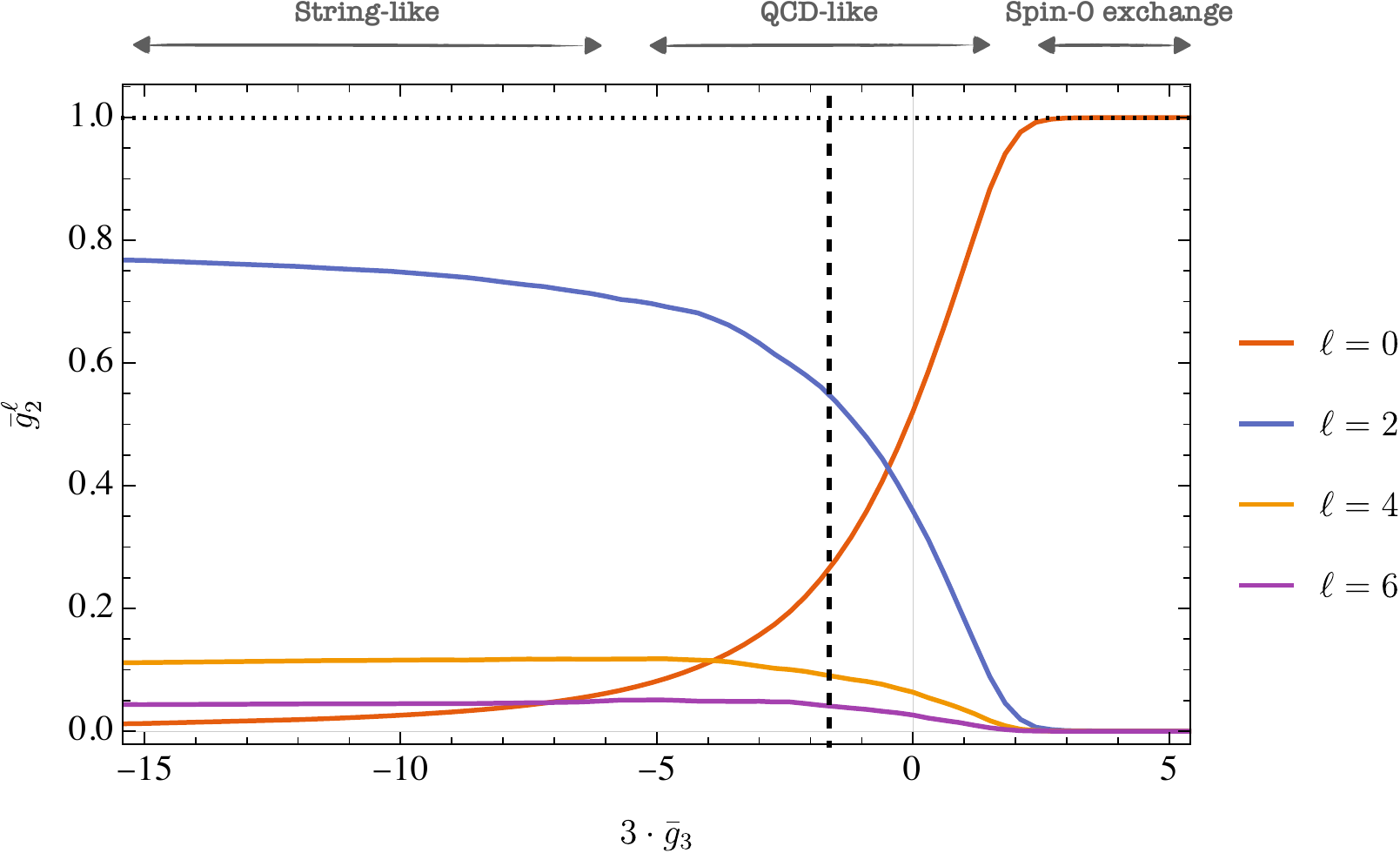}
	\caption{The numerical values of the branching ratio of the cross section $\bar g_2^\ell$ defined in \eqref{eq:coefficients_g2l} of the amplitudes on the lower boundary in figure \ref{fig:bound} as a function of $\bar g_3$. The colors represent different values of the angular momentum $\ell=0,2,4,6$.}
	\label{fig:sim-rules}
\end{figure}

\subsection{Phenomenology of the boundary}
At each point on the boundary of the allowed region, given by the orange line in figure \ref{fig:bound}, there is a unique amplitude which we can reconstruct numerically. The amplitudes on the boundary are called extremal. Our goal in this subsection is to understand the physics contained in these extremal amplitudes.
 
The simplest way to analyze a scattering process is to plot partial amplitudes $\cT_\ell(s)$ (also called partial waves) for several values of angular momentum $\ell=0,2,4,\ldots$. Recall, that in the physical region the scattering amplitudes in $d=4$ space-time dimensions is written in terms of the partial amplitudes as
\begin{equation}
	\label{eq:decomposition}
	\cT(s,t) = 2\pi \sum_{\ell=0,2,\dots}(1+2\ell)P_\ell\left(1+ \frac{2t}{s}\right)\,\cT_\ell(s),
\end{equation}
where $P_\ell(x)$ is the Legendre polynomial. For more details, see appendix \ref{app:NL_unitarity}.

In order to diagnose how the various degrees of freedom in the amplitude distribute in the spin channels, it is useful to look at the spin decomposition of some observables. A convenient choice is given by $g_2$ and its sum-rule representation, derived in appendix~\ref{app:sum-rules}. One has
\begin{equation}
g_2=\frac{1}{\pi}\int_0^\infty ds \frac{\Im\cT(s,0)}{s^3}= \frac{1}{\pi}\int_0^\infty ds \frac{\sigma_{\text{tot}}}{s^2} \geq 0,
\end{equation}
where $\sigma_\text{tot}$ is the total cross section.
If we decompose the amplitude in the forward direction in partial waves inside the integral, we can define the following dimensionless coefficients that can be thought as \emph{branching ratios} of the cross section in each spin channel~\cite{EliasMiro:2022xaa},
\begin{equation}
	\label{eq:coefficients_g2l}
	\bar g_2^{\ell} \equiv 2g_2^{-1}\,(1+2\ell)\int_0^\infty ds \frac{\Im\cT_\ell(s)}{s^3}\geq 0\,.
\end{equation}
The values $\bar g_2^{\ell}$, interpreted as a sort of partial cross sections, are thus a good indication of the spin content of the amplitude. 
In figure \ref{fig:sim-rules} we present the value of $\bar g_2^{\ell}$ using different colors for different spins as a function of $\bar g_3$ that we use to parametrize the boundary.
The sum rule \eqref {eq:coefficients_g2l} is already well approximated by the sum of the branching ratios $\bar g_2^{\ell}$ with few angular momenta $\ell=0,2,4,6$. By looking at the figure we see that when $\bar g_3\gg 1$ the $\ell=0$ ratio gives the most important contribution to the cross-section, suggesting that our bounds might have a simple description in terms of a perturbative QFT amplitude. On the other hand, when $\bar g_3\ll -1$, the sum of the higher spin ratios dominate and we might envision a string-like amplitude. Close to the minimum value of $\bar g_4$ where we do expect strongly coupled physics, the spins arrange in a non-trivial way.

\begin{figure}[t]
	\centering
	\includegraphics[scale=0.5]{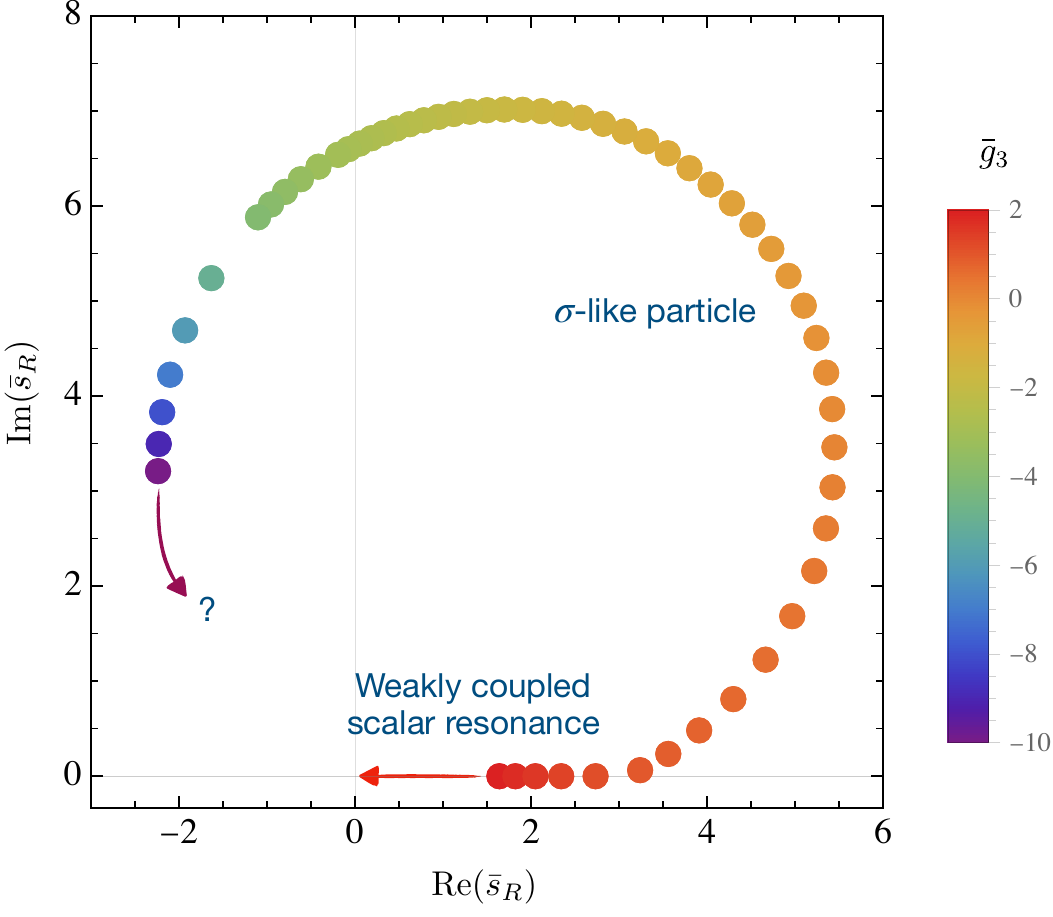}
	\caption{Trajectory of the lowest lying scalar zero as function of $\bar g_3$. }
	\label{fig:followZeroAlongg3}
\end{figure}

We can refine our physical picture by studying the position of the spectrum of resonances as a function of $\bar g_3$. Consider partial waves
\begin{equation}
	\mathcal{S}_\ell(s)\equiv1+\frac{i}{8} \mathcal{T}_\ell(s).
\end{equation}
Resonances are given by zeros $s_R$ of the partial waves $\mathcal{S}_\ell(s_R)=0$ in the upper half plane.
Their position in the complex plane can be interpreted in terms of the mass and the width $s_R=(m+i\Gamma/2)^2$ of the unstable particle \cite{Guerrieri:2022sod}. 
In figure~\ref{fig:followZeroAlongg3} we plot the position of the lowest lying resonance in the $\ell=0$ partial wave for several values of $\bar g_3$.
Connecting the dots we obtain a smooth continuous trajectory parametrized by $\bar g_3$.
The \emph{nature} of this zero changes as we follow it along the trajectory . When $\bar g_3\gg 1$, in red, the zero shows up close to the real axis with a mass parametrically larger than its width $m\gg \Gamma$. Then it can be described as a light weakly coupled scalar particle. 
On the opposite extreme, when $\bar g_3\ll -1$, its interpretation is obscure as it goes close to the left cut region $s<0$. Around the minimum value of $\bar g_4$, when $\bar g_3\sim -1$, its real and imaginary part are of the same order and we can interpret it as a \emph{$\sigma$-like particle}, as the famous scalar resonance in QCD.

Combining the information in figure \ref{fig:sim-rules} and figure \ref{fig:followZeroAlongg3}, we can envision the existence of three distinct regions which will be investigated below,
\begin{align}
	&\text{Region I  (\emph{QCD-like}}):\qquad\qquad\qquad\;\;  \,|\bar g_3| \sim 0,\\
	&\text{Region II (\emph{string-like}}):\qquad\qquad\qquad\,\bar g_3 \ll -1,\\
	&\text{Region III  (\emph{spin-0 exchange}}):\qquad\,\,\,\,\,\,\, \bar g_3 \gg +1.
\end{align}

\subsection{Complex spin analysis of the amplitude minimizing $\bar g_4$}

\begin{figure}
	\centering
	\includegraphics[scale=0.65]{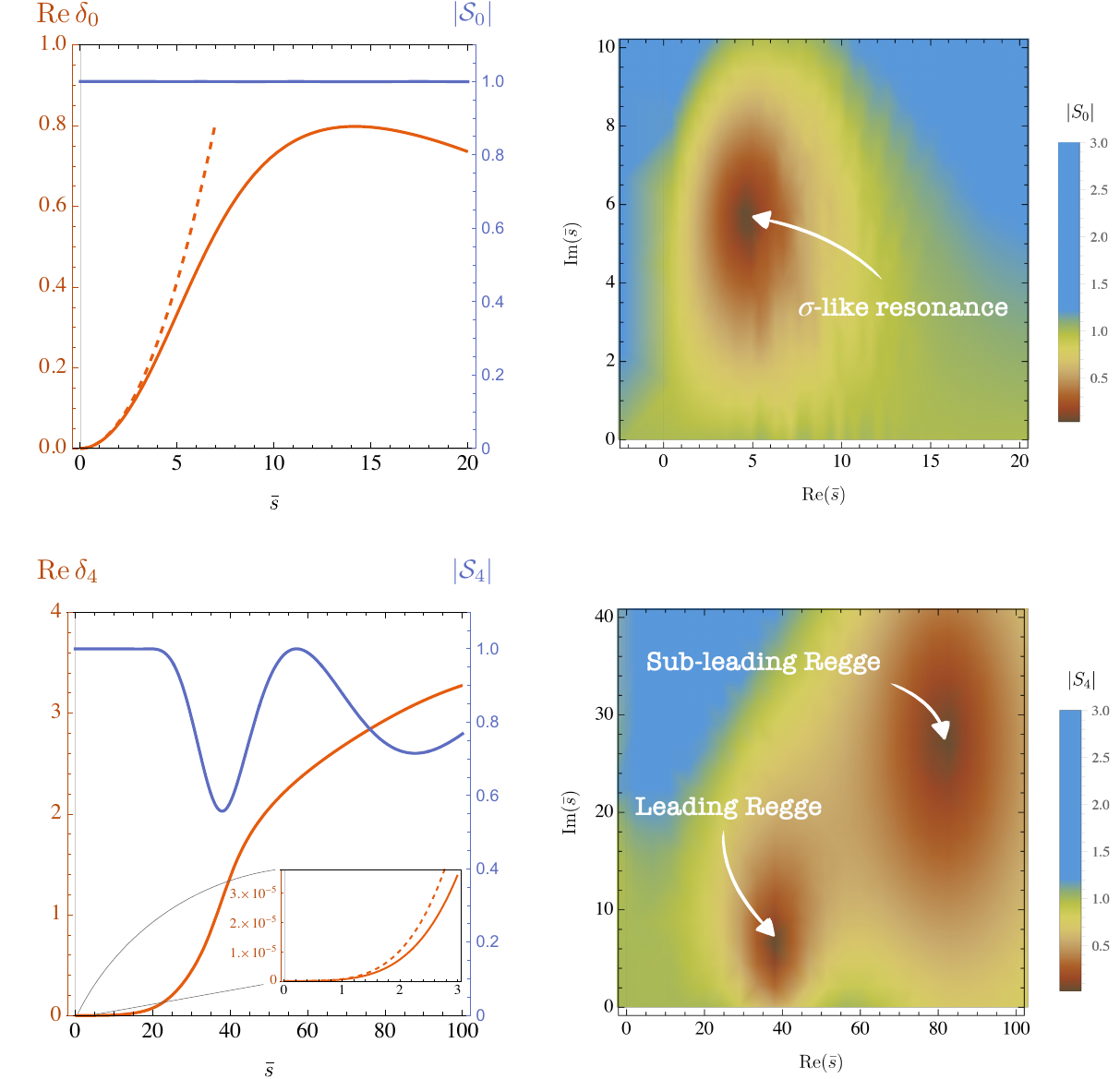}
	\caption{On the left, we plot the phase shifts for both spins $\ell=0,4$ in red, and the absolute value $|\mathcal{S}_\ell|$ of the corresponding partial waves in blue. Red dashed is the one-loop EFT approximation expected to be reliable up to the scale $\bar s\sim1$. On the right, we plot the absolute value $|\mathcal{S}_\ell|$ in the complex $\bar s$ plane. There we observe the presence of zeros that we interpret as resonances. }
	\label{fig:ComplexPlaneMing4}
\end{figure}

In Region I, based on the previous analysis, we do expect to find amplitudes resembling QCD. As a benchmark point we take the minimum value of $\bar g_4$. We study the spectrum of unstable resonances contained in the amplitude.

Consider the phase shift $\delta_\ell(s)$ defined via
\begin{equation}
	S_\ell (s) = e^{2i \delta_\ell(s)}.
\end{equation}
The typical signature of a weakly coupled resonance is a rotation of the phase by $\pi$. Its mass can be estimated by looking at the energy at which the phase passes through $\pi/2$. 
The rotation of the phase can be associated to the presence of a zero of the partial wave in the complex plane close to the real axis.
When the resonance is not weakly coupled it is more difficult to detect it by the phase shift analysis. The better way to study resonances is to inspect the complex plane as we do below.

In figure~\ref{fig:ComplexPlaneMing4} on the top panels we study $\delta_0(s)$ and $|S_0(s)|$ in the complex $\bar s$ plane. We observe the presence of a broad resonance far away from the real axis. We call it $\sigma$ resonance.
In figure~\ref{fig:ComplexPlaneMing4} on the bottom panels, we plot the phase shift $\delta_4(s)$ and the absolute value of the spin four partial wave $|S_4(s)|$. In this case, the $\delta_4(s)$ clearly passes through $\pi/2$ around $\bar s\sim 40$, and it keeps growing at higher energies. 
In the complex plane we find two zeros: one close to the real axis, and a heavier one with large imaginary part. 
The former is interpreted as a weakly coupled spin four particle, the latter as a spin four $\sigma$-like resonance.
Unitarity saturation in the spin four channel is not yet achieved by our numerics as it can be appreciated by looking at the solid blue line in the bottom-left panel. We know empirically that the lack of convergence does not affect the position of the resonances, but rather its width (the imaginary part) -- see  also the discussions in~\cite{Guerrieri:2018uew, Bose:2020cod, Guerrieri:2022sod}. 

The theory of complex angular momenta~\cite{Gribov:2003nw} -- see also~\cite{Correia:2020xtr} for a recent review -- suggests that resonances with higher spins are different realizations of the same object, the \emph{Reggeon}. Reggeons have a mass that continuously depend on the spin-$\ell$ parameter. Projecting the amplitude on various real spin channels we expect to follow the resonance along its trajectory, the famous \emph{Regge trajectory}. The analytic continuation in spin of partial waves is performed through the Froissart-Gribov formula
\begin{equation}
S_\ell(s)=1+\frac{i}{32 \pi}\int_0^\infty dt \frac{8}{\pi s}Q_\ell\left(1+\frac{2t}{s}\right)\text{Disc}_t \mathcal{T}(s,t),
\end{equation}
where $Q_\ell(x)$ is the Legendre polynomial of the second kind.

\begin{figure}
	\centering
	\includegraphics[width=1\linewidth]{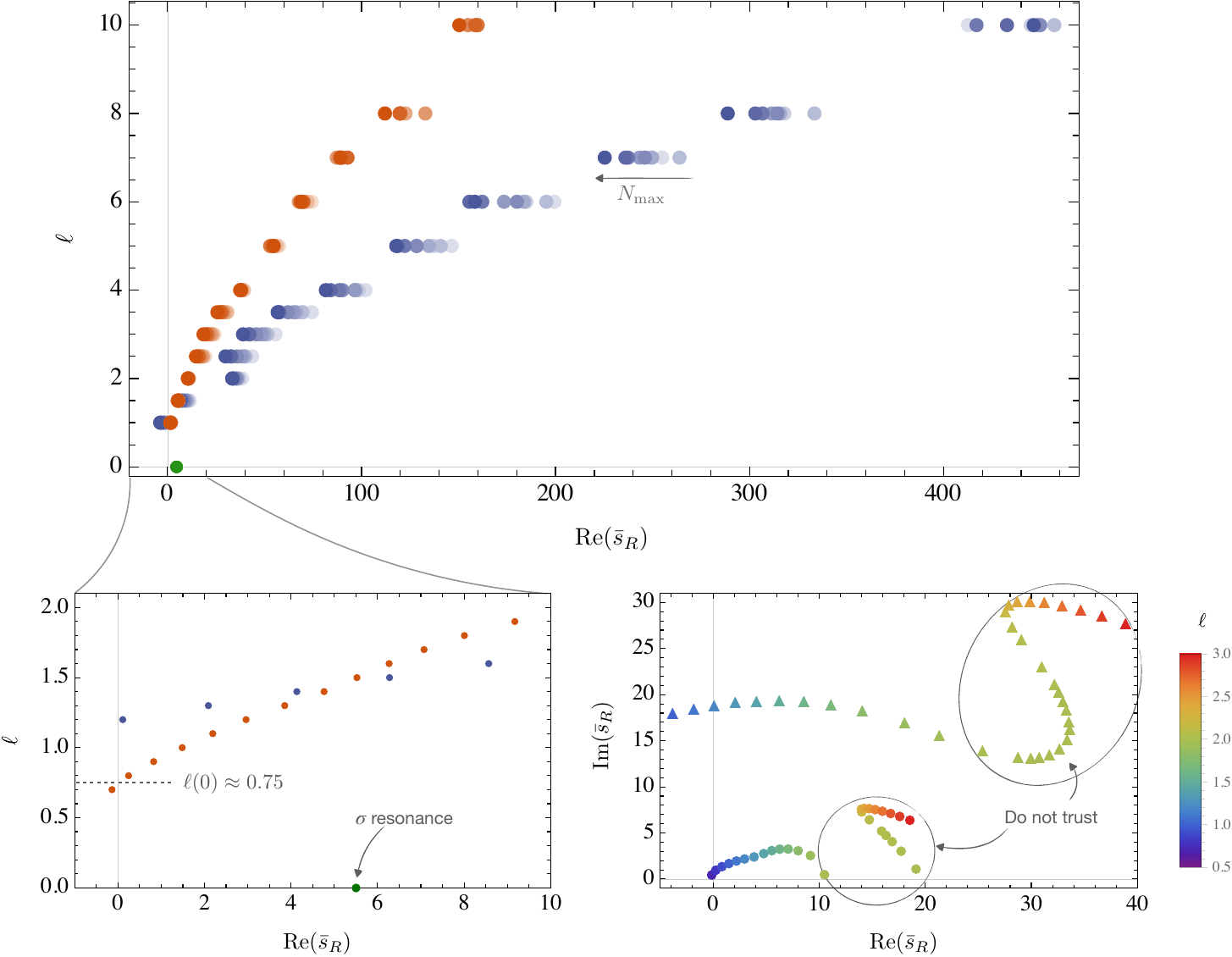}
	\caption{Chew-Frautschi diagram for the amplitude that minimize $\bar g_4$.  On the top panel, the leading (in red) and sub-leading (in blue) trajectories are presented up to spin 10. The spin 0 (green) resonance is not part of a trajectory. In this plot, we show various $N_{\text{max}}$ to express the convergence of the position of the zero and we use darker color for larger $N_{\text{max}}$. On the bottom left panel, we zoom on the trajectories at low spin. The leading (red) is well converged in this region, this is not the case for the sub-leading (blue) and we expect it to move up .  On the bottom right panel, we present the position of the resonance in the complex plane for both the leading (circle) and sub-leading (triangle) trajectory. In the circled region, we do not trust the position of the resonance.}
	\label{fig:reggeTrajectory_ming4}
\end{figure}

We collect the various resonances in the Chew-Frautschi diagram for different spins $\ell\leq 10$ in figure~\ref{fig:reggeTrajectory_ming4} (top panel). The resulting distribution is suggestive. We clearly see two trajectories.\footnote{Notice, that the Regge trajectories have been recently found numerically also in \cite{Bose:2020cod}, see figure 12 and in \cite{Guerrieri:2022sod}, see figure 4.} The red dots belong to the \emph{leading trajectory}, since it contains the particles that have the lowest mass for each spin; the blue dots to the \emph{sub-leading trajectory} or more informally the \emph{daughter trajectory}. The resonances in the leading trajectory are weakly coupled, and the linear universal behaviour emerges in our data \cite{Caron-Huot:2016icg} as we follow it to higher spins. The daughter particles are $\sigma$-like and deep in the complex plane, therefore the linearity is lost. In figure~\ref{fig:reggeTrajectory_ming4} (bottom left) we zoom in the low spin region where $\ell\leq 2$. We observe the crossing of the two trajectories around spin $\ell_c\sim 1.4$. We indicate the lightest $\sigma$-like resonance of $\ell=0$ studied in figure~\ref{fig:followZeroAlongg3} with a green dot. In the zoomed figure we appreciate how it does not belong to any trajectory, but rather is an isolated particle. This is in agreement with the fact that the analytic continuation in spin of our ansatz only converges for $\Re \ell>\ell_0 =0$.

In figure~\ref{fig:reggeTrajectory_ming4} (bottom right), we collect the position of the resonance in the complex plane. On the bottom left of the figure, we see how the leading trajectory moves and similarly for the sub-leading on the top part of the plot. In the complex plane $\bar s$ it becomes clear that the crossing of the two trajectories observed above is just the result of a projection.
Moreover, we find a surprising behaviour between spins $2\lessapprox\ell\lessapprox3$. The corresponding arcs are highlighted with `circles'. We believe the behaviour there is not physical, but rather consequence of poor convergence in that region of the complex-spin plane. We remind the reader that our numerics are affected by a systematic error due to truncation both in the size of the ansatz $N_{\text{max}}$ and the number of spin constraints imposed. The amplitude studied in this section gives the best approximation of the minimum value of $\bar g_4$ at hand, but it is not necessarily well converged for all energies and spins. Indeed, the position of the resonance in those regions still has a strong dependence on $N_\text{max}$.

\begin{figure}
	\centering
	\includegraphics[scale=1]{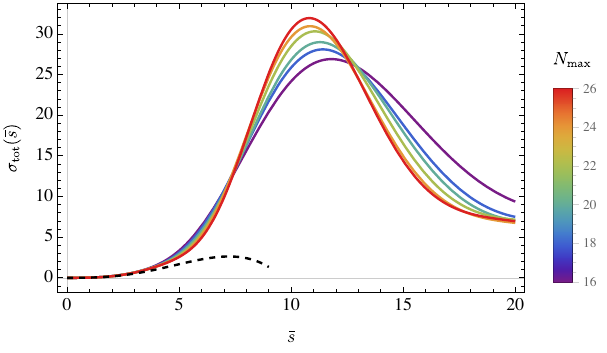}
	\caption{Cross section $\sigma_{\text{tot}} =s^{-1}\text{Im}_s \mathcal{T} (s,t=0) $ for the amplitude that minimize $\bar g_4$. The dashed black line is the EFT expansion in \eqref{eq:crossSectionEFT}. The different colors is the cross section of the amplitude with different $N_{\text{max}}$.  }
	\label{fig:crossSectionming4}
\end{figure}

Analytic continuation of unitarity for complex spin is rigorously possible only in the elastic region. When the scattering is elastic it is possible to prove that $|S_\ell|=1$ for $\Re \ell >\ell_0$. For the scattering of massless particles there is no such statement. However, since the extremal amplitudes we study are mostly elastic at low energies and for all spins, it would be natural to expect that $|\mathcal{S}_\ell|\approx 1$. In appendix~\ref{sec:complexSpinSl}, figure~\ref{fig:complexSpinSl} we plot the absolute value of $\mathcal{S}_\ell$ on the real axis for for various spins in the neighboring of $\Re \bar s_R$. We check that our expectations are realized when the Regge trajectories are smooth and monothonic,  but fail for the problematic spins $2\lessapprox\ell\lessapprox3$, the circled arcs in figure~\ref{fig:reggeTrajectory_ming4}.

We call $\ell(t)$ the parametric curve that describes the Regge trajectory as function of $t$. The highest intercept $\ell(0)$ among all Regge trajectories determines the asymptotic behaviour of the total cross section
\begin{equation}\label{eq:reggeBehavior}
\sigma_{\text{tot}}= \frac{\text{Im}\mathcal{T}(s,0)}{s}\sim s^{\ell(0)-1}
\end{equation}
From figure~\ref{fig:reggeTrajectory_ming4}, we can extract the intercept of the leading Regge trajectory to be $\ell(0)\approx0.75$. However, due to the crossing phenomenon at spin $\ell_c\approx 1.4$ the daughter trajectory becomes dominant at higher energies. 
Even though the numerical convergence is harder for the daughter trajectory, we do estimate that the leading intercept will have $\ell(0)\sim 1$, a feature typical of the pomeron trajectory.\footnote{As shown on figure~\ref{fig:reggeTrajectory_ming4}, the position (real part) of the leading (red) trajectory is well converged in $N_{\text{max}}$, this is not true for the sub-leading. Thus, it  is  difficult to determine the intercept of this trajectory precisely and similarly for the position of the crossing between the two trajectories.}$^,$\footnote{
There is a tension, though, between this Regge analysis, and the behaviour of our ansatz for the amplitude at infinity. The former predicts an almost constant total cross-section, the latter a vanishing cross-section at infinity. Numerical evidence suggests that resonances of higher spins appear as we increase the size of the ansatz used in our numerics, and therefore we expect to solve this tension only asymptotically. We leave the detailed investigation of this conundrum to further study.}

Finally, we look at the total cross section as a function of $\bar s$ in figure~\ref{fig:crossSectionming4}. Curves with different colours correspond to different values of $N_\text{max}$. At low energies we can compare the non-perturbative cross section with the EFT approximation in dashed black
\begin{equation}\label{eq:crossSectionEFT}
\sigma^{\text{EFT}}_\text{tot}=s^{-1}\text{Im}_s \mathcal{T}^{\text{EFT}}(s,t=0)=\frac{7}{80 \pi}g_2^2 s^3+\frac{1}{60\pi} g_2 g_3 s^4.
\end{equation}
In this case, since the UV physics is strongly coupled, the cutoff of the EFT approximation is given by naive dimensional analysis and is expected to be around $\bar s\sim 1$. However, when $\bar s<1$ the EFT approximation is good for all angles, and this explains the nice agreement of the phase shifts in figure~\ref{fig:ComplexPlaneMing4}, where the EFT approximation is plotted in dashed red. The peak in the total cross-section happens at the mass of the lightest spin two resonance.

One could repeat a similar analysis to all amplitudes in the Region I. By simple inspection, we can generically say that all amplitudes in this region share the same qualitative features, although the details of the spectrum change.

\subsection{Asymptotic bounds and tree level physics}
\label{sec:aymtptotic_bound}

When we take $\bar g_3$ large, we expect the EFT approximation in equation \eqref{eq:soft_1} to be valid in the regime when
\begin{equation}
|\bar g_3| \bar s^3\ll \bar s^2 \implies \bar s \ll |\bar g_3|^{-1}.
\label{eq:asymptoticEFT}
\end{equation}
To retain the relevant terms when $\bar s \ll |\bar g_3|^{-1}$, it is convenient to define the rescaled variable $\hat s= \bar s |\bar g_3|^{-1}$. In this new regime, the amplitude can be approximated by\footnote{Similar ideas have been developed and applied to the branon $S$-matrix of confining strings~\cite{to_appear_axions}.}
\begin{equation}
\mathcal{T}=\frac{1}{\bar g_3^2}\left((\hat s^2+\hat t^2+\hat u^2)+\hat s \hat t \hat u + \frac{\bar g_4}{\bar g_3^2}(\hat s^2+\hat t^2+\hat u^2)^2+\mathcal{O}(\hat s^5) \right)+\mathcal{O}(\bar g_3^{-4}),
\label{eq:tree_expansion}
\end{equation}
where the logs are suppressed by higher powers of $\bar g_3$.
The large $\bar g_3$ parameter plays the role of $\hbar^{-1}$ in a putative effective action expansion, and the the amplitude is well approximated by the expansion of a tree level UV theory. 
Following this logic, we might expect the coefficient of the $\mathcal{O}(\hat s^4)$ term in~\eqref{eq:tree_expansion} to be an order one number. The only possibility is that $\bar g_4\sim \bar g_3^2$, and this prediction is precisely satisfied by the asymptotic behaviour of the boundary region in~\ref{fig:bound_equivalent}.

In the $\bar g_3\to \infty$ limit, we can apply the technology developed in \cite{Caron-Huot:2020cmc} 
to determine bounds on the ratio $ \tfrac{\bar g_3}{\sqrt{\bar g_4}}$. We obtain\footnote{This bound is obtained using 35 null constraints, ie. $n=16$ in the notation of~\cite{Caron-Huot:2020cmc}, see for example figure 10 there. }
\begin{equation}
-15.67\lessapprox  \frac{\bar g_3}{\sqrt{\bar g_4}}\leq 3\sqrt{2}\simeq 4.2
\end{equation}
which is compatible with our asymptotic numerical estimate \eqref{eq:bound_ratio}. 
The upper bound is saturated by a simple tree level massive scalar exchange  \cite{Caron-Huot:2020cmc}, compatible with the spin zero dominance observed in figure~\ref{fig:sim-rules}, and the presence of a light weakly coupled scalar resonance in figure~\ref{fig:followZeroAlongg3}.
The difference in the lower bound is likely due to the difficulty of the primal numerical algorithm to converge at large negative $g_3$, therefore we expect the gap between the two methods to close. 
For the lower bound, there is no known function that saturates it.

\begin{figure}
	\centering
	\includegraphics[scale=1]{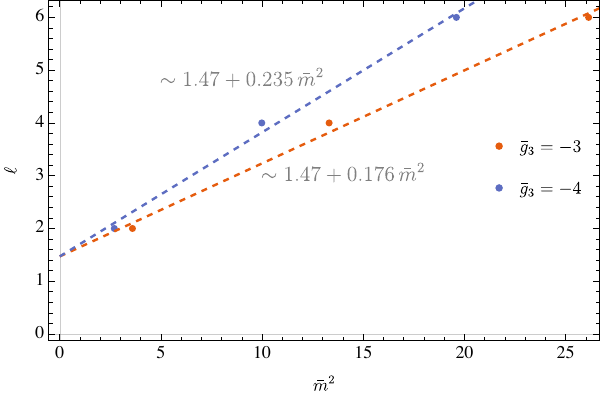}
	\caption{The masses of the lightest resonances plotted versus their spin. They lie on a line. We interpret this line as a Regge trajectory. }
	\label{fig:regge}
\end{figure}

Studying the extremal amplitudes that saturate the asymptotic bounds we can guess the tree level UV completion in the large $\bar g_3$ limit. Moreover, using some simple unitarization model, we can estimate the behaviour of the amplitude away from the tree-level approximation for large, but finite $\bar g_3$. We report this analysis in the next paragraphs.

\begin{figure}[t]
	\centering
	\includegraphics[scale=0.88]{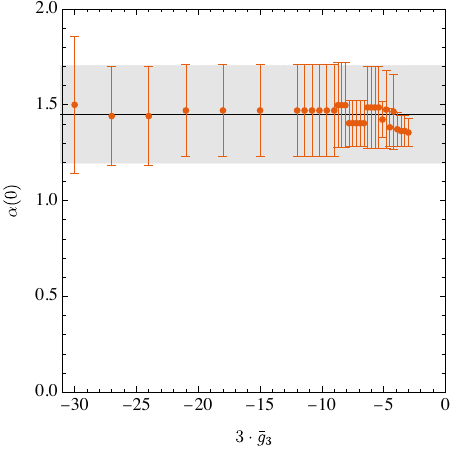}\hspace{0.5cm}
	\includegraphics[scale=0.9]{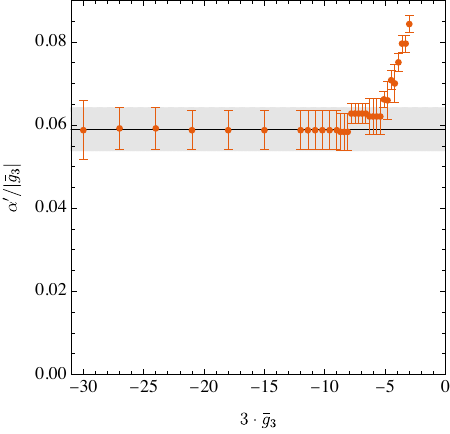}
	\caption{Estimation of the coefficients $\alpha(0)$ and $ \alpha'$ of the Regge trajectories defined in \eqref{eq:Regge_trajectory} for different values of $\bar g_3$.}
	\label{fig:RegionA_alphaprime}
\end{figure}

\paragraph{Region II (\emph{string-like})}\mbox{}\\
In region II, the amplitude receives contribution from all spins starting at spin $\ell =2$ according to figure \ref{fig:sim-rules}. Performing a partial wave analysis we observe the presence of weakly coupled massive resonances for all spins $\ell\geq 2$. We estimate the mass of these resonances and produce the corresponding Chew-Frautschi plot of the spin $\ell$ of the particle as a function of its mass squared $\bar m^2\equiv\sqrt{g_2}m^2$ for each value of $\bar g_3\ll -1$.\footnote{The resonance are weakly coupled and are close to the real axis and therefore $\Re(\bar s_R)\approx\bar m^2$.} In figure \ref{fig:regge} we show the resonances for $\ell\leq 6$ for two values of $\bar g_3$. All the points approximately lie on a single straight line, and we fit it with an ansatz of the form 
\begin{equation}
	\label{eq:Regge_trajectory}
	\ell (\bar m^2 ) \approx \alpha(0) + \alpha'\times \bar m^2.
\end{equation}
The coefficient $\alpha(0)$ is called the Regge intercept and $\alpha '$ the slope. The estimation of the coefficients $\alpha(0)$ and $ \alpha'$ of the Regge trajectories for different values of $\bar g_3$ is presented in figure \ref{fig:RegionA_alphaprime}. 
To match the tree level EFT description in \eqref{eq:tree_expansion}, all mass parameters in the amplitude must scale as $\bar g_3^{-1}$ at leading order.
Indeed, we experimentally observe that\footnote{We analyze here the leading Regge trajectory. The amplitude are not converged enough to extract the daughter trajectories.} 
\begin{equation}
	\label{eq:Regge_trajectory_coefficients}
	\alpha(0) = 1.45\pm 0.25 ,\qquad \alpha'/|\bar g_3|= 0.059\pm 0.005
\end{equation}

We can conclude that in this region,  the amplitudes exhibit a string-like behavior.  In the limit $\bar g_3\to -\infty$, the  amplitude at the boundary is described by a meromorphic string-like amplitude with the following properties 
\begin{itemize}
	\item Spin $\ell$ contribution given by the limit $\bar g_3 \ll -1$ of figure \ref{fig:sim-rules}. In particular, it has no spin $0$ exchange.
	\item Leading Regge trajectory with intercept and slope given by \eqref{eq:Regge_trajectory_coefficients}
	\item Ratios of Wilson coefficients given by $ \frac{\bar g_3}{\sqrt{\bar g_4}} \approx -15.67$.
\end{itemize}
We are not aware of any string model which generates amplitudes compatible with these properties.  For the recent works on string amplitudes, see \cite{Caron-Huot:2016icg,Cheung:2022mkw,Cheung:2023adk,Cheung:2023uwn}. 

\begin{figure}[t]
	\centering
	\includegraphics[scale=1]{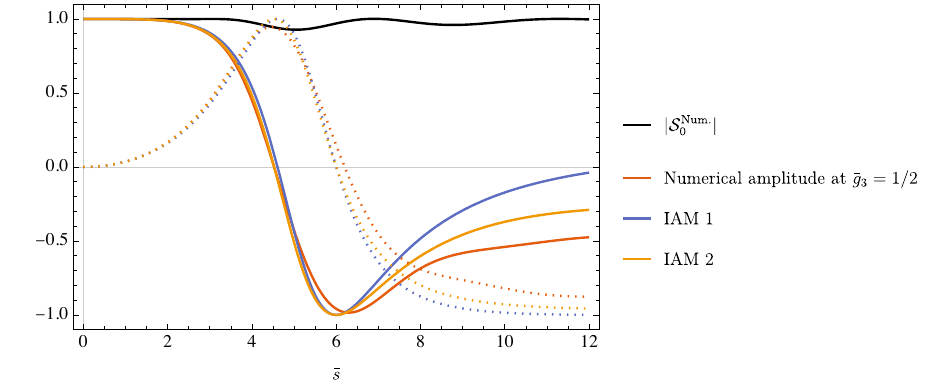}
	\caption{Real and imaginary part of the spin zero partial amplitude $\mathcal{S}_0$, obtained on the boundary of figure \ref{fig:bound} at $\bar g_3 =1/2$. Our numerical result is depicted in red. The blue and orange lines provide the rough estimates obtained using the inverse amplitude method reviewed in appendix \ref{app:IAP}.}
	\label{fig:spin0_IAMvsNum}
\end{figure}

\paragraph{Region III  (\emph{spin-0 exchange})}\mbox{}\\
In region III, the boundary precisely approaches the tree level positivity prediction
\begin{equation}
\label{eq:limit_scalar_bound}
\lim_{\bar g_3\to \infty}\frac{\bar g_3}{\sqrt{\bar g_4}}= 3\sqrt{2}.
\end{equation}
The positivity bound is saturated by a simple tree-level scalar exchange
\begin{equation}
	\label{eq:tree_amplitude_main}
	\mathcal{T}_\text{spin 0}(s,t,u) \equiv -\lambda ^2
	\left(\frac{m^2}{s-m^2}+\frac{m^2}{t-m^2}+\frac{m^2}{u-m^2}+3\right),
\end{equation}
where $\lambda$ is the dimensionless coupling and $m$ the mass of the scalar exchanged. Expanding this amplitude around $s=0$ keeping $x\equiv \cos\theta$ fixed we obtain an expansion of the form \eqref{eq:tree_expansion}. Matching the two expressions, upon identifying $s$ in~\eqref{eq:tree_amplitude_main} with $\hat s$, we can express the low energy Wilson coefficients $g_2, g_3$ in terms of the UV parameters $\lambda, m$
\begin{equation}
	\label{eq:values_gn_tree}
	g_2=\frac{\lambda^2}{m^4}, \qquad
	g_3=\frac{3\lambda^2}{m^{6}}.
\end{equation}
The matching predicts also $g_4=\frac{\lambda^2}{2m^{8}}$, which is consistent with~\eqref{eq:limit_scalar_bound}.
In addition to the correct asymptotic behavior of the bound, this statement is supported by two additional observation. First, in this region, the sum-rule \eqref{eq:coefficients_g2l} is dominated by the spin-0 contribution as shown in figure \ref{fig:sim-rules}. Second, in figure \ref{fig:spin0_IAMvsNum} we plot the real and imaginary part of the spin zero partial amplitude extracted on the boundary of figure \ref{fig:bound} at $\bar g_3 =1/2$. Our result is depicted in red. In blue and orange we provide an estimate of this partial amplitude obtained by applying the Inverse Amplitude Method (IAM) to \eqref{eq:tree_amplitude_main}, and using the value for the mass and the coupling given by the matching conditions in~\eqref{eq:values_gn_tree} for $\bar g_3=1/2$. 
We review the IAM in appendix \ref{app:IAP}. The agreement is good in the region around the light resonance. At higher energies the amplitude is not trivial but contains broad higher spin resonance. Their mass does not scale with $\bar g_3$, so in units of the mass of the light resonance $\hat s$, they decouple.

\section{S-matrix machinery for massless particles and its improvements}
\label{sec:machinery_scenario_1}

In this section we introduce and explain our improvements of the numerical $S$-matrix bootstrap machinery needed for studying massless particles. We start in subsection \ref{sec:review} by reviewing the standard machinery. Our improvements are then presented in subsections \ref{sec:improvement_1} and \ref{sec:improvement_2}.

\subsection{Machinery review}
\label{sec:review}
In order to study scattering amplitudes numerically in the {\it generic scenario} we use the machinery introduced in \cite{Paulos:2017fhb} combined with an extension of this machinery introduced in \cite{Guerrieri:2020bto} needed for working with massless particle. It consists of writing the following ansatz
\begin{equation}
	\label{eq:ansatz_1}
\text{Ansatz 1: }	\quad	\mathcal{T}_\text{ansatz}(s,t,u) = \sum_{a,b,c}
	\alpha_{abc}\, \rho^a(s)\rho^b(t)\rho^c(u) + \mathbb{N}(s,t,u).
\end{equation}
where $\alpha_{abc}$ are real coefficients. Crossing requires that they are fully symmetric in their indices. Due to the relation among the Mandelstam variables $s+t+u=0$, not all terms are independent in the expansion \eqref{eq:ansatz_1}, it is standard to require 
\begin{equation}
	\forall (abc)\neq 0:\qquad
	\alpha_{abc} = 0\,,
\end{equation}
which removes the redundant terms. 
The $\rho$-variable is defined as
\begin{equation}
	\label{eq:rho_definition}
	\rho(z) \equiv \frac{\sqrt{4m^2-z_0}-\sqrt{4m^2-z}}{\sqrt{4m^2-z_0}+\sqrt{4m^2-z}},
\end{equation}
where $m$ is the mass of the external particle (in our case $m=0$) and $z_0$ is an arbitrary parameter, numerical results do not depend on it. We choose $z_0=-1$. The infinite sum in \eqref{eq:ansatz_1} is truncated in practice in such a way that $\max(a+b+c)\leq N_{\text{max}}$. The numerics is ran for several values $N_{\text{max}}$. This allows for an $N_{\text{max}}\rightarrow\infty$ extrapolation.

The first term in the ansatz \eqref{eq:ansatz_1} has the square root branch point at $s=0$, however the effective amplitude of Goldstones given by \eqref{eq:soft_1} -  \eqref{eq:L_expression} has a log branch-point because of the log-terms in \eqref{eq:L_expression}. The latter can never be represented by the former at finite $N_{\text{max}}$. In order to deal with this issue we have added the additional term $ \mathbb{N}(s,t,u)$ in \eqref{eq:ansatz_1}. Let us explain how to obtain its explicit expression.
Let us first denote the non-analytic part of the effective amplitude \eqref{eq:soft_1} -  \eqref{eq:L_expression} (which contains the log-terms) by
\begin{multline}
	\label{eq:N_definition}
	N(s,t,u) \equiv - \frac{g_2^2}{480\pi^2} \Big(
	s^2 (41 s^2+t^2+u^2) \log\left(-s\sqrt{g_2} \right) +\\
	t^2 (s^2+41 t^2+u^2) \log\left(-t\sqrt{g_2} \right) +
	u^2 (s^2+t^2+41 u^2) \log\left(-u\sqrt{g_2} \right) \Big).
\end{multline}
We can then also define
\begin{equation}
	\label{eq:loga_ansatz}
	\mathbb{N}(s,t,u) \equiv - \frac{g_2^2s_0^{4}}{480\pi^2}\left[\chi_s^2\, \mathbb{H}(s|t,u) \log\chi_s + \chi_t^2 \, \mathbb{H}(t|s,u) \log\chi_t+\chi_u^2 \, \mathbb{H}(u|t,s)\log\chi_u\right],
\end{equation}
where we the $\chi$-variable is defined as
\begin{equation}
	\label{eq:chi_def}
	\chi_s \equiv \frac{1}{4}(\rho_s-1)^2 - \frac{1}{4}(\rho_s-1)^3 = \frac{s}{s_0}-3\left(\frac{s}{s_0}\right)^2+ \cO(s^{5/2})
\end{equation}
and the $\mathbb{H}$-function is given by
\begin{equation}
	\label{eq:H_standard}
	\mathbb{H}(s|,t,u) = 41 \chi_s^2+ \chi_t^2 + \chi_u^2.
\end{equation}
 The non-analytic expression \eqref{eq:N_definition} has an unlimited growth at large energies and, thus, violates unitarity. The term \eqref{eq:loga_ansatz} instead remains finite at large energies. It is constructed in such a way that at low energies it reproduces \eqref{eq:N_definition}, namely
\begin{equation}
		\mathbb{N}(s,t,u) = N(s,t,u)-
	s^2\left[\beta_{1,1} s^2+\beta_{1,2} tu+ \beta_{1,3} (t^2 +u^2 )\right]\log(-s_0\sqrt{g_2}) + \cO(s^5).
\end{equation}

Next, we require that at low energies the ansatz \eqref{eq:ansatz_1} precisely reproduces the effective amplitude \eqref{eq:soft_1}. Performing the expansion of \eqref{eq:ansatz_1} we simply get
\begin{multline}
	\label{eq:expansion}
    \mathcal{T}(s,t,u) =N(s,t,u)+
	\alpha_{000} + k_1 (\alpha, x) s^{1/2} + k_2 (\alpha, x) s + k_3 (\alpha, x) s^{3/2} + k_4 (\alpha, x) s^{2}\\
	+ k_5 (\alpha, x) s^{5/2} + k_6 (\alpha, x) s^{3} + k_7 (\alpha, x) s^{7/2}  + k_8 (\alpha, x) s^{8}  + O(s^{9/2}),
\end{multline}
where $k_i(\alpha, x)$ are some functions which depend on the coefficients of the ansatz and $x\equiv \cos\theta$, where $\theta$ is the scattering angle. These functions can be easily obtained in Mathematica, their form, however, is too large (and depends on $N_\text{max}$) in order to be written here.
Comparing this expansion with \eqref{eq:soft_1}  -  \eqref{eq:L_expression} we conclude that
\begin{multline}
	\alpha_{000} = 0,\quad
	k_1= 0,\quad
	k_2= 0,\quad
	k_3= 0,\quad
	k_4= g_2,\\
	k_5= 0,\quad
	k_6= g_3,\quad
	k_7= 0,\quad
	k_8= g_4.
\end{multline}
After solving these constraints and plugging the solution back to \eqref{eq:ansatz_1} we obtain the ansatz which is fully compatible with the effective amplitude \eqref{eq:soft_1}  -  \eqref{eq:L_expression}.

The ansatz \eqref{eq:ansatz_1} satisfies crossing and maximal analyticity by construction. The non-linear unitarity formulated in the semi-definite positive way \eqref{eq:unitarity_practice} is imposed numerically using SDPB \cite{Simmons-Duffin:2015qma,Landry:2019qug}.

The parameter $g_2$ defines the mass scale. In practice for performing the numerics we set it to some constant value, for example one could chose $g_2=1$. Our results do not depend on this choice since we always work with dimensionless quantities. It can happen, however, that at finite $N_\text{max}$ there is a preferred value of $g_2$ for which the numerics converges faster. In table \ref{tab:Optimalg2Foring4} we show the lower bound on $\bar g_4$ (keeping $\bar g_3$ free) for different choices of $g_2$. We see that the best bound is achieved for roughly $g_2=500$.
When we construct bounds on $\bar g_4$ as a function of $\bar g_3$ the optimal value of $g_2$ changes with $\bar g_3$. As a result one needs to carefully adjust it in order to achieve the most optimal numerical convergence.
\begin{table}[t]
	\centering
	\begin{tabular}{c|cccccc}
		\toprule
		$g_2$&$1$&$10$&$100$&$500$&$1000$&$10000$\\
		\midrule
		$\bar g_4$&$0.01978$&$0.01245$&$0.01100$&$0.01062$&$0.01070$&$0.01171$ \\
		\bottomrule
	\end{tabular}
	\caption{Lower bound on $\bar g_4$ for different choices of $g_2$. The best bound is achieved at $g_2=500$. Here we use $N_{\text{max}}=20$ and $ L_{\text{max}}=80$.}
	\label{tab:Optimalg2Foring4}
\end{table}

\subsection{Improvement 1}
\label{sec:improvement_1}

Let us consider the first derivative in $t$ of the amplitude in the forward limit, namely 
\begin{equation}
	\label{eq:dert_forward}
	\partial_t\mathcal{T}(s,t=0,u).
\end{equation}
Up to $O(s^5)$, \eqref{eq:dert_forward} is finite in the series representation \eqref{eq:soft_1}. 
It is reasonable to expect that  \eqref{eq:dert_forward} is finite at any order but we do not know of any general proof.
From now on we assume that \eqref{eq:dert_forward} is finite. This for instance allows to write the following sum-rule for the $g_3$ coefficient
\begin{equation}
	\label{eq:g3_sum-rule}
	g_3 = \frac{2}{\pi}  \int_0^\infty ds \left(\frac{3\Im \cT(s,t)}{2s^4} -\frac{\partial_t\Im \cT(s,t)}{s^3}\right)_{t\to0}.
\end{equation}
See appendix \ref{app:sum-rules} for its derivation.

Consider now the ansazt \eqref{eq:ansatz_1}. Evaluating \eqref{eq:dert_forward} using this ansatz we immediately see that  \eqref{eq:dert_forward} diverges. This is an unpleasant behaviour which prevents us for instance from using the sum-rule \eqref{eq:g3_sum-rule}. In what follows we introduce a minor modification of the ansatz \eqref{eq:ansatz_1} which removes this divergence.

Let us define the following auxiliary variable
\begin{equation}
	\zeta(z) \equiv \frac{1}{4}(\rho(z)-1)^2 = -\frac{z}{z_0} + \mathcal{O}(z^{3/2}).
\end{equation}
Using it we can then define the following modified $\myRho$-variable
\begin{equation}
	\begin{aligned}\label{eq:newRho}
		\myRho^0(z) &\equiv 1,\\
		\myRho^a(z) &\equiv \zeta(z)\rho^a(z),\quad a>1.
	\end{aligned}
\end{equation}
Using this variable we propose the new ansatz
\begin{equation}
	\label{eq:ansatz_2}
\text{Ansatz 2: }	\quad \mathcal{T}_\text{ansatz}(s,t,u) = \sum_{a,b,c}
	\alpha_{abc}\, \myRho^a(s)\myRho^b(t)\myRho^c(u) + \mathbb{N}(s,t,u).
\end{equation}
One can explicitly check that the ansatz \eqref{eq:ansatz_2} has finite value of \eqref{eq:dert_forward}. Below, we will check that this modification does not change the results of the bounds, see for example figure~\ref{fig:ming4both} for comparison of extrapolated bound. 
\begin{figure}[t]
	\centering
	\includegraphics[scale=1]{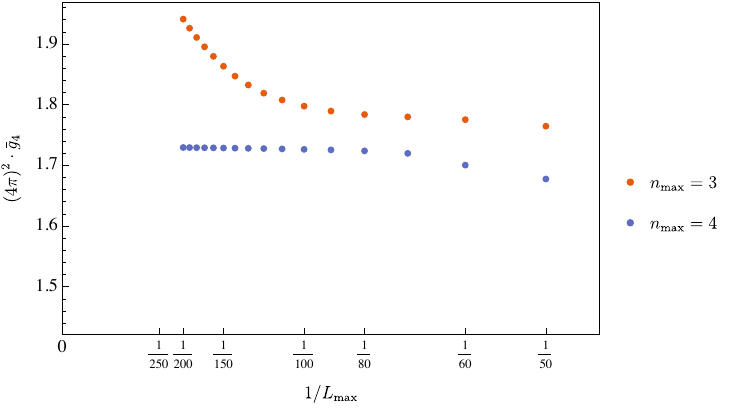}
	\caption{The lower bound on $\bar g_4$ using the ansatz 2.
		Both curves are constructed using the ansatz  \eqref{eq:ansatz_2} with \eqref{eq:eq:H_new}. In red curve with  $n_{\text{max}}=3$, for large values of $L_{\text{max}}$ it starts changing rapidly. We refer to this region as the ramp.  For the red curve, the ramp disappears and we are left with the stable bound. The bounds here are constructed with $N_{max}=18$.}
	\label{fig:Nmax18Ramp}
\end{figure}

\subsection{Improvement 2}
\label{sec:improvement_2}

In order to numerically impose the unitarity condition \eqref{eq:unitarity_practice} we need to project scattering amplitudes to partial amplitudes. The latter are labelled by the angular momentum $\ell=0,2,4,\ldots$. In theory we need to impose \eqref{eq:unitarity_practice} for infinitely many values $\ell$. In practice, however, this is impossible, and we impose \eqref{eq:unitarity_practice} for a finite set of angular momenta $\ell=0,2,4,\ldots, L_\text{max}$.

The parameter $L_\text{max}$ is the new player in the setup. In order to obtain physical results one needs to find the $L_\text{max}$ value high enough such that the numerical output stabilizes and remains independent of the change of $L_\text{max}$.  For some numerical problems (especially for massless particles) the output keeps changing with $L_\text{max}$. In those situations one runs the numerics for several different values of $L_\text{max}$ and then performs the extrapolation $L_\text{max}\rightarrow \infty$.

In \cite{Haring:2022sdp} it was observed that the numerics breaks down for large values of $L_\text{max}$. This was a severe obstacle for performing the $L_\text{max}\rightarrow \infty$ extrapolation in \cite{Haring:2022sdp} reliably. In this subsection we reproduce the same phenomenon in the case of neutral Goldstones using the ansatz \eqref{eq:ansatz_2} and propose a modified ansatz which  solves this problem. 

Recall, that the ansatz \eqref{eq:ansatz_2} contains the log-terms via the expressions \eqref{eq:N_definition} and \eqref{eq:H_standard}. We identified the ramp behaviour with the expression \eqref{eq:H_standard}. It turned out that \eqref{eq:H_standard} is too restrictive. In order to add more freedom we can instead write
\begin{equation}
	\label{eq:eq:H_new}
	\mathbb{H}(s|t,u)  = \sum_{a,b,c=0}^{n_\text{max}} \gamma_{abc}\, \chi_s^a\chi_t^b \chi_u^c,
\end{equation}
where $n_\text{max}$ is a truncation parameter for the sum. The real coefficients $\gamma_{abc}$ are symmetric in the last two indices as required by crossing. We tune these coefficients in such a way that in the low energy expansion we get
\begin{equation}
	\mathbb{H}(s|t,u) = 41 s^2+t^2+u^2 + O(s^3).
\end{equation}
This ensures that the ansatz correctly reproduces the effective amplitude \eqref{eq:soft_1} at low energies. Due to this requirement \eqref{eq:H_standard} is the same as \eqref{eq:eq:H_new} but only up to $O(s^3)$ terms.

\begin{figure}[t]
	\centering
	\includegraphics[scale=0.8]{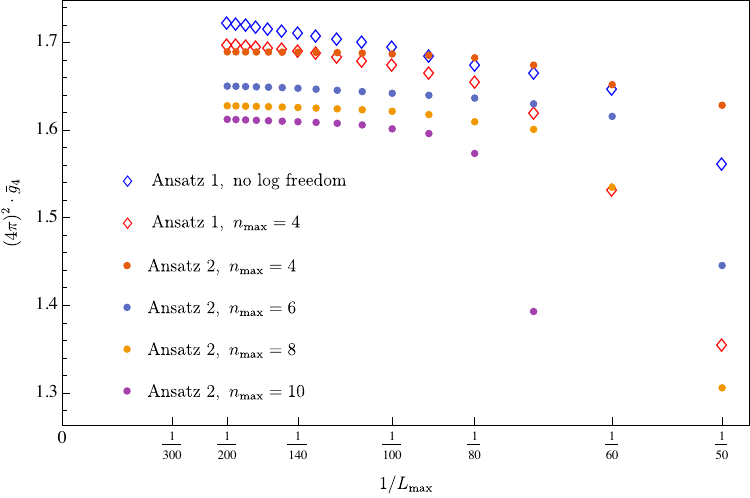}
	\caption{Lower bound on $\bar g_4$ for the two types of the ansatz \eqref{eq:ansatz_1} and \eqref{eq:ansatz_2} and various values of $n_\text{max}$ in \eqref{eq:eq:H_new}. Here we focus only on $N_\text{max}=20$.}
	\label{fig:CompareSpinExtrapolation}
	
	\vspace{1.5cm}
	
	\includegraphics[scale=0.9]{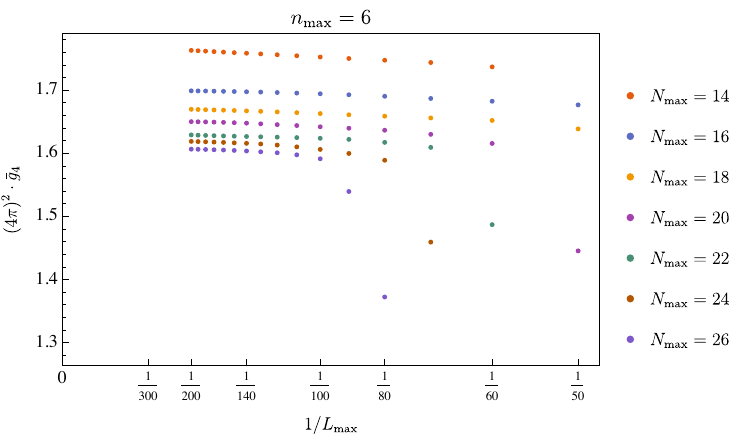}
	\caption{Lower bound on $\bar g_4$ as a function of $L_\text{max}$ for various values of $N_\text{max}$. Here we use the ansatz 2 and keep $n_\text{max}=6$.}
	\label{fig:exampleExtrapolation}
\end{figure}

The main result is given in figure \ref{fig:Nmax18Ramp}. In this figure we compute the lower bound on $\bar g_4$ (keeping $\bar g_3$ free) as a function of $L_\text{max}$ using the ansatz \eqref{eq:ansatz_2} with a different amount of freedom in the $log$ term $\mathbb{N}$. We observe that for low $n_{\text{max}}$ the bound becomes abnormally strong. Increasing $L_\text{max}$  further we observe that the numerics  breaks down (becomes unfeasible). This is clearly an unphysical and problematic behavior. The rapid increase of the red line will be referred to as the ramp.\footnote{For the ansatz \eqref{eq:ansatz_1} the ramp starts for higher values of $L_\text{max}$ and, thus, is much less pronounced in the given interval of $L_\text{max}$.} As we increase $n_{\text{max}}$, the ramp is ``pushed away'' to higher   $L_\text{max}$ and the extrapolation  to infinite spin can be taken. For moderate values of $L_\text{max}$ the two curves are consistent with each other, however, for large $L_\text{max}$ the blue curve remains stable and does not exhibit any ramp behavior. Notice, that as the two curves are built with different $n_{\text{max}}$, the freedom in ansatz used for the blue curve is bigger which explains why they do not overlap at intermediate spin.  In the limit $N_\text{max}\rightarrow\infty$ this difference completely disappears as will be shown in the end of this subsection.

 \begin{figure}[t]
	\centering
	\begin{subfigure}{0.60\textwidth}
		\includegraphics[width=\linewidth]{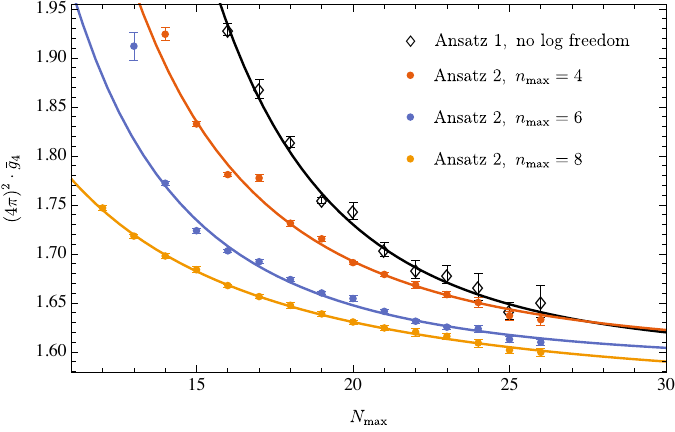}
		\caption{Extrapolation in Nmax}
		\label{fig:ming4vsNmax}
	\end{subfigure}
	\begin{subfigure}{0.38\textwidth}
		\includegraphics[width=\linewidth]{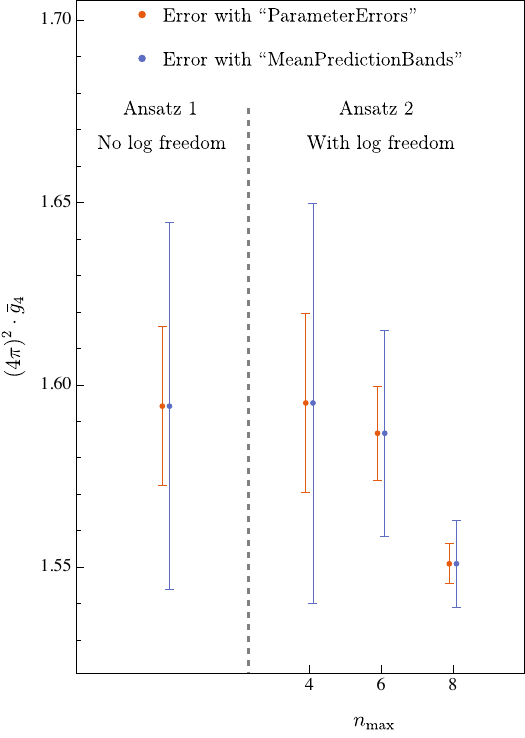}
		\caption{Result}
		\label{fig:ming4Extrapolated}
	\end{subfigure}
	\caption{Extrapolation of the lower bound on $\bar g_4$ in different scenarios. In the left plot we show the numerical data by dots and the extrapolation by the solid line. In the right plot we show the extrapolated bound for different cases. Solid lines show the error bars. The results are fully consistent among each other within the error bars.}
	\label{fig:ming4both}
\end{figure}

Let us now focus on the case of $N_\text{max}=20$ and compare bounds with several different values of $n_\text{max}$.  We do it in figure \ref{fig:CompareSpinExtrapolation}.
We can see that the ramp disappears in all these cases. In figure \ref{fig:exampleExtrapolation} we fix $n_\text{max}=4$ and scan instead over various values of $N_\text{max}$. We see that the behaviour of the bound remains stable with $L_\text{max}$  for any $N_\text{max}$ (no ramp).  Using the data presented in figure \ref{fig:exampleExtrapolation} we can perform the $L_{\text{max}}\rightarrow \infty$ extrapolation. We use several different fits in order to estimate the extrapolated value and its error. In the repository with our numerical data linked to this paper we provide further details about the fits used. In figure \ref{fig:ming4vsNmax} we presented our $L_{\text{max}}\rightarrow \infty$ extrapolated bound (with error bars) on $\bar g_4$ as a function of $N_\text{max}$. 

We have introduced several types of ansatz which at finite $N_\text{max}$ lead to consistent, but slightly different answers. Let us now perform the $N_\text{max}\rightarrow \infty$ extrapolation and show that different choices lead to the same result within the error bars. We will focus again on the $\bar g_4$ minimization problem (keeping the $\bar g_3$ value free). Our extrapolation for various ansatz choices is presented in figure \ref{fig:ming4vsNmax}. The $\bar g_4$  bound in the $N_\text{max}\rightarrow \infty$ limit for various ansatz choices is summarized in particular in figure \ref{fig:ming4Extrapolated}.

\section{Model-dependent bounds}
\label{sec:modeldepBound}
Let us now study the allowed space of amplitudes in the scenario in which we have an EFT-inspired model. We impose that the effective field theory approximation of the amplitude is valid in the extended region $s\in[0, M^2]$ for all angles. This is imposed by the condition \eqref{eq:condition_2}. In practice we impose non-linear unitarity in the whole range of energies $s\in [0,\infty]$. Thus, we can impose the condition \eqref{eq:condition_2} only on the imaginary part of the amplitude (the real part will be adjusted by the numerics accordingly), namely
\begin{equation}
	\label{eq:condition_2c}
	s\in[0,M^2]:\qquad
	| \text{Im}_s \mathcal{T}(s,\cos\theta) - \text{Im}_s \mathcal{T}_\text{EFT}(s,\cos\theta) |\leq
	\text{err}(s,\cos\theta).
\end{equation}
We emphasize here that the model is  defined by  a `band'  given by $\text{Im}_s \mathcal{T}_\text{EFT}(s,\cos\theta)\pm	\text{err}(s,\cos\theta)$ constraining the amplitude up to a cut-off scale $M^2$. For example, experimental measure of the differential cross-section could be used to define the model. In that follow, we will describe how this can be implemented numerically in subsection~\ref{sec:machinery_scenario_2} and present the result for a simple model in subsection~\ref{subsec:scenario_2}.

\subsection{Machinery for the model-dependent bounds}
\label{sec:machinery_scenario_2}
Here we explain how to construct numerically scattering amplitudes which obey crossing, maximal analyticity and unitarity at all energies and satisfy the condition \eqref{eq:condition_2c} describing a \emph{model}. Since we have two regions with $s\leq M^2$ and $s>M^2$ it is natural to write the ansatz as a sum of two terms
\begin{equation}
	\label{eq:ansatz_PB}
	\mathcal{T}_\text{ansatz} (s,t,u) = \mathcal{T}^1_\text{ansatz} (s,t,u) + \mathcal{T}^2_\text{ansatz} (s,t,u).
\end{equation}
Here the first term takes care of the low energy behaviour of the amplitude, in other words it must approximately describe \eqref{eq:soft_1} in the extended region $s\in[0, M^2]$. The second term takes care of the high energy behaviour and it is adjusted by the numerics in such a way that the full ansatz \eqref{eq:ansatz_PB} obeys non-linear unitarity. Our proposal for the two terms in \eqref{eq:ansatz_PB} reads as\footnote{Note that $\mathcal{T}^1_\text{ansatz}$ is the same as the ansatz in the universal bounds in \eqref{eq:ansatz_2}.}
\begin{align}
	\label{eq:ansatz_PB_1}
	\mathcal{T}^1_\text{ansatz} (s,t,u)  &\equiv 
	\mathbb{N}(s,t,u) + \sum_{a,b,c=0}^{
		N_\text{max}^1}\alpha_{abc}\,\myRho_1^a(s)\myRho_1^b(t)\myRho_1^c(u),\\
	\mathcal{T}^2_\text{ansatz} (s,t,u)  &\equiv 
	\sum_{a,b,c=0}^{N^2_\text{max}}\beta_{abc}\, \rho_2^a(s)\rho_2^b(t)\rho_2^c(u),
\end{align}
where $\mathbb{N}(s,t,u)$ is defined in \eqref{eq:loga_ansatz} and the two types of the $\rho$-variable are defined as
\begin{equation}
	\rho_1(z) \equiv \frac{\sqrt{-z_0}-\sqrt{-z}}{\sqrt{-z_0}+\sqrt{-z}},\qquad
	\rho_2(z) \equiv \frac{\sqrt{M^2-z_0'}-\sqrt{M^2-z}}{\sqrt{M^2-z_0'}+\sqrt{M^2-z}}\,,
\end{equation}
with $\myRho_1$ defined by \eqref{eq:newRho} using $\rho_1$.
Here $z_0<0$ and $z'_0\leq0$ are real negative parameters which can be chosen at our will. In practice we choose $z_0=-1$ and $z_0'=0$. The above variables are constructed in such a way that they contain a branch cut starting from $s=0$ and $s=M^2$ respectively. 

In order to impose unitarity \eqref{eq:unitarity} we choose a set of points using the Chebyshev grid. In practice we pick 100 points in the $s\in [0,M^2]$ region and 200 points in the $s>M^2$ region. We impose non-linear unitarity at these 300 points and for angular momenta $\ell=0,2,\ldots, 150$.

In the region $s\in [0,M^2]$ on top of the non-linear unitarity we also impose the condition \eqref{eq:condition_2c}. In practice this is done as follows. We pick a linear grid in the $x\equiv\cos\theta$ variable and use 10 points in the interval $x\in[0,1]$.\footnote{For a $t-u$ symmetric amplitude, the amplitude is symmetric in $x\to -x$ and we could only consider the half interval  $x\in[0,1]$. In general, one could choose a grid in the full interval  $x\in[-1,+1]$. } We impose \eqref{eq:condition_2c}  for all possible combinations of 100 points in $s$ (distributed with the Chebyshev grid) and the 10 points in $x$.\footnote{In practice, we observe that convergence in the size of the grid in $x$ is fast and 10 points were a safe choice.} On top of it we also project the condition \eqref{eq:condition_2c} into partial wave. We impose this condition only for spin-$\ell$ which lead to non-zero projection of the `band' and the precise number of spin depend on the model. We impose the projected condition \eqref{eq:condition_2c} for 100 points in $s$ and each $\ell$.

We observe that for the runs we have performed, one can choose a fixed size of ansatz $\cT^1_{\text{ansatz}}$ \eqref{eq:ansatz_PB_1} needed to reproduce the desired low energy behavior. Then an extrapolation can be performed in $N^2_{\text{max}}$ which dictate the freedom of the high energy behavior.  In practice, we chose $N^1_{\text{max}}=10,\, n_{\text{max}}=8$ and converge in $N^2_{\text{max}}$.  

\begin{figure}[t]
	\centering
	\includegraphics[scale=1]{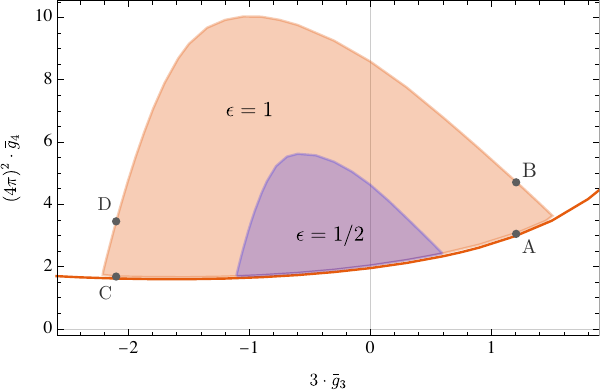}
	\caption{The allowed region of parameters is enclosed inside the islands. Two colors represent two different values of the parameter $\epsilon$ introduced in \eqref{eq:err_function_choice}. The two values we use are $\epsilon=1/2$ and $\epsilon=1$. For building this plot we also chose $\xi=1$ defined in \eqref{eq:g2_M_relation}. The orange line indicate the bound of figure \ref{fig:bound}. In figure \ref{fig:EFT_UVCompletion}, we plotted the imaginary part of the amplitude at the points A,B,C,D.}
	\label{fig:EFT_bounds}
	\vspace{2cm}
		\includegraphics[scale=0.75]{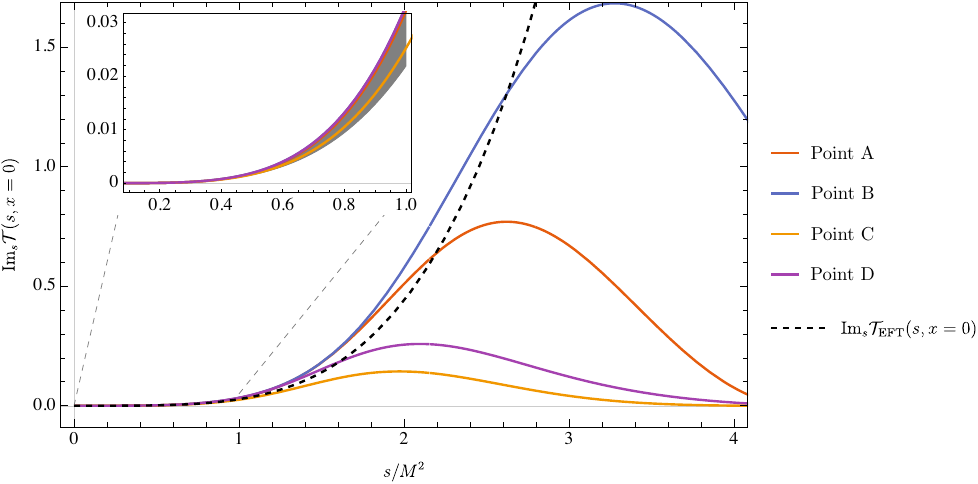}
	\caption{Examples of UV completion of the amplitude at different point labeled by A,B,C,D in the boundary of figure \ref{fig:EFT_bounds} for $\epsilon=1$. We picked pairs of points close to each extremity and plotted the imaginary part of the amplitude in the forward limit at the upper/lower boundary. In gray, we represented the band chosen using  \eqref{eq:choice_Center_band} and \eqref{eq:err_function_choice}.}
	\label{fig:EFT_UVCompletion}
\end{figure}

\subsection{Result using a model for the IR amplitude}\label{subsec:scenario_2}
Here we define a simple model based on the EFT expansion of the amplitude. The error function can be defined as follows. In appendix \ref{app:one-loop_unitarity} we compute the  imaginary part of $\mathcal{T}^{(5)}_\text{EFT}$ term exactly. We can equate the error function to this term. However, in order to take into account corrections due to higher order terms $\mathcal{T}^{(n)}_\text{EFT}$ with $n\geq 6$ we replace $\bar g_3$ by $\epsilon$ and use $\epsilon$ as an extra parameter. We find then the `band' of the model by
\begin{align}
    \text{Im}_s \mathcal{T}_\text{EFT}(s,t,u)&= \frac{g_2^2}{480 \pi} s^2 (41 s^2 + t^2 + u^2)\label{eq:choice_Center_band}\\
	\text{err}(s,t,u)&=\frac{\epsilon}{1920 \pi}\,g_2^{5/2}s^3\big(36s^2-4(t^2+u^2)\big)	\label{eq:err_function_choice}\, .
\end{align}
This model is thus defined by two dimensionless  parameters, namely $\xi$ defined in \eqref{eq:g2_M_relation} and $\epsilon$ defined in \eqref{eq:err_function_choice}. In what follows we will focus our attention on the $\xi=1$ case. This choice favors the class of amplitudes with strongly coupled UV completions. See the discussion below \eqref{eq:g2_M_relation} for the explanation.
Regarding $\epsilon$, we impose the condition \eqref{eq:condition_2c} for two different values, namely $\epsilon=1/2$ and $\epsilon=1$. The resulting bound is given in figure \ref{fig:EFT_bounds}. As one can see the lower bound agrees with the universal bound in figure~\ref{fig:bound_normal}.
Indeed, the lower inequality of the extended EFT condition in~\eqref{eq:condition_2c} is never active, in agreement with the fact that the amplitudes in Region I admit a strongly coupled UV completion, and that in some sense are extremal.

It is interesting to see how the EFT amplitude is UV completed along the boundary. In figure \ref{fig:EFT_UVCompletion}, we show the imaginary part of the amplitude in the forward limit at two pairs ot points for the case $\epsilon=1$. For $s<M^2$, the constraint imposed \eqref{eq:condition_2c} is clearly satisfied and is saturated along the upper boundary. In contrast, for $s>M^2$, the imaginary part is unitarised differently depending on the point on the boundary.

\section{Conclusions and outlook}
\label{sec:conclusions}

In this paper we initiated the exploration of the space of $2\to 2$ scattering amplitudes of massless neutral Goldstone bosons in four dimensions. We numerically determined the allowed region of the first two non-universal Wilson coefficients appearing in the $2\to 2$ scattering, dubbed $\bar g_3$ and $\bar g_4$ both in the \emph{universal scenario}, and by imposing a low energy model for the amplitude.

In the {\it universal scenario} we used the numerical approach of \cite{Paulos:2017fhb,Guerrieri:2020bto} where an analytic ansatz is written for the amplitude and unitarity is imposed numerically. In this paper we proposed two new improvements of the ansatz in section \ref{sec:machinery_scenario_1}. The first improvement allows us to use the usual Bootstrap ansatz to match the fixed-$s$, small $t$ behaviour of the amplitude predicted by the EFT expansion in~\eqref{eq:soft_2_3}. The second removes the \emph{ramp} instability for large number of angular momentum constraints $L_\text{max}$ discovered in \cite{Haring:2022sdp}.
We presented the main result in the {\it universal scenario} in figure \ref{fig:bound}. 

On the boundary given by the orange line in figure \ref{fig:bound} we can reconstruct the amplitudes numerically. 
Analyzing the branching ratios of the cross section in various spin channels and the behaviour of the lightest spin zero resonance, we identified three different regions along the boundary. The two asymptotic regions correspond respectively to weakly coupled string amplitudes with approximately linear Regge trajectories, and weakly coupled scalar exchanges. In between these two regimes, the amplitude resembles a strongly coupled theory like QCD.
For the special point that minimizes $\bar g_4$, we performed for the first time a thorough analysis of the spectrum of resonances continuing the scattering amplitude to complex spins. We could interpolate the first two Regge trajectories by following the resonances in spin, and extract reliably their intercepts without fitting. We observed a curious crossing phenomenon among the two trajectories.

The amplitudes we extracted on the boundary for large values of $\bar g_3$ and $\bar g_4$ are not fully converged for the finite size ansatz we used. This is rather unsatisfactory. In the future, one could attempt to solve this issue by centering the $\rho$-variables at multiple different points. One of the realizations of this idea called the \emph{wavelet ansatz} was advocated for in \cite{EliasMiro:2022xaa}. Numerical convergence is controlled also by the number of spin constraints that should be taken to be very large. 
However, computing the partial wave projection with high precision and for very large spin is numerically expensive. In~\cite{Guerrieri:2022sod}, the authors introduced the notion of \emph{unitarity in the sky}, and showed how dramatically improves the convergence in the spin cutoff. We believe that by combining the improvements introduced in this paper with both the ideas mentioned in this paragraph might pave the way to more efficient numerical studies of massless scattering amplitudes.

The universal bounds determined in this paper apply to the dilatons of $\mathcal{N}=4$ SYM on the Coulomb branch. It would be interesting to consider also the other massless Goldstones of the $R$-symmetry breaking, and study the mixed system of dilatons and Goldstones. Another possible direction is to generalize the Bootstrap of neutral Goldstones to other dimensions. In three dimensions, we could study the low energy dynamics of $\mathcal{M}_2$ branes, using maximal supersymmetry to obtain the bounds on the Wilson coefficients from the scattering of neutral massless scalars.
On the other hand, in higher dimensions such bounds could be used to constrain the Wilson coefficients of the effective action of various strings compactifications and to test the consistency of the available non-perturbative results with the principles of analyticity, crossing, and unitarity.

In the {\it model-dependent scenario} we assumed that the amplitude is described by an EFT-inspired model at energies below a chosen  
cut-off scale $M$. In section \ref{sec:machinery_scenario_2} we proposed a concrete numerical implementation of this idea inspired by \cite{Chen:2021pgx}. The main result in the {\it model-dependent scenario} is given in figure \ref{fig:EFT_bounds}.
This method is not limited to massless particles. One could use this approach also for the scattering of physical pions, and improve the results obtained in~\cite{Guerrieri:2018uew} by injecting the available experimental or lattice data as low energy constraints into the bootstrap setup.

As briefly reviewed in the end of appendix \ref{app:absence_bounds} the Wilson coefficients of the dilaton scattering in six dimesions is related to the difference of the UV and IR $a$-anomalies denoted by $\Delta a$. In general, as it is well known, there are no bounds on $\Delta a$ in 6d from the $2\to 2$ scattering of dilatons. However, adding an extra assumption/knowledge about their IR behaviour up to some energy scale one can apply our {\it model-dependent scenario} to get some bounds on $\Delta a$. In QFTs with explicit conformal symmetry breaking one can introduce a massless scalar probe field (also often called the dilaton) which carries information about $\Delta a$. No new bounds can be constructed on $\Delta a$ in this case using our technique.

\section*{Acknowledgements}
We thank Miguel Correia, Aditya Hebbar, Sasha Monin, Francesco Riva, Biswajit Sahoo and Alexandre Zhiboedov  for useful discussions.
We also thank Joan Elias-Mir\'o for discussions and comments on the draft.
The work of FA is supported by the Galileo Galilei Institute Boost Fellowship.
Research at the Perimeter Institute is supported in part by the Government of Canada through
NSERC and by the Province of Ontario through MRI.
AG is supported by the European Union -
NextGenerationEU, under the programme Seal of Excellence@UNIPD, project acronym CluEs. 
The work of KH is supported by the Simons Foundation grant 488649 (Simons Collaboration on the Non-perturbative Bootstrap) and by the Swiss National Science Foundation through the project 200020 197160 and through the National Centre of Competence in Research SwissMAP.
DK also thanks the organisers and participants of the ``S-matrix Bootstrap Workshop V'' where part of this work was presented.
DK is supported by the SNSF Ambizione grant PZ00P2\_193411. 
\appendix

\section{Non-linear unitarity}
\label{app:NL_unitarity}
The partial amplitudes for the scattering of identical particles in a general number of dimensions have the following form
\begin{equation}
	\mathcal{S}_\ell(s) = 1 + \frac{i}{\mathcal{N}_d(s)} \mathcal{T}_\ell(s),  
\end{equation}
where the interacting part of the partial amplitudes are defined as
\begin{equation}\label{eq:PartialWaveDef}
	\mathcal{T}_\ell(s) \equiv \int_{-1}^{+1} dx \,\mu_{d,\ell}(x) \mathcal{T}(s,t(x),u(x)).
\end{equation}
Here $\ell=0,2,4,\ldots$ is the angular momentum and $x\equiv \cos\theta$ is the cosine of the scattering angle. The object $\mathcal{N}_d(s)$ for massless particles is given by
\begin{equation}
	\label{eq:factor_N}
	\mathcal{N}_d(s) \equiv 2^{d-1}s^{\frac{4-d}{2}},
\end{equation}
and the measure reads as
\begin{equation}
	\label{eq:measure}
	\mu_{d,\,\ell}(x) =  \frac{j!\,\Gamma\left(\frac{d-3}{2}\right)}{4\,\pi^{(d-1)/2}\Gamma(d-3+\ell)}
	\times
	(1-x^2)^{\frac{d-4}{2}} C_\ell^{(d-3)/2}(x).
\end{equation}
The relation between the $(t,u)$ and $(s,x)$ variables for massless particles is given by
\begin{equation}
	\label{eq:tu_expressions}
	t = -\frac{s}{2}\,(1-x),\quad
	u = -\frac{s}{2}\,(1+x).
\end{equation}

The relation \eqref{eq:PartialWaveDef} can be inverted. As a result the interacting part of the scattering amplitude can be written as a sum of interacting partial amplitudes as follows
\begin{equation}
	\label{eq:decomposition_partial_amplitudes}
	\cT(s,t) = \sum_{\ell=0,2,\dots} c_\ell^{(d)} P_\ell^{(d)}\left(1+ \frac{2t}{s}\right)\,\cT_\ell(s),
\end{equation}
where the $d$-dimensional Legendre polynomials are defined as
\begin{equation}
	\label{eq:definition_P}
	P^{(d)}_\ell(x) \equiv {}_2F_1 \left(-\ell,\ell+d-3,\frac{d-2}{2},\frac{1-x}{2} \right).
\end{equation}
By using the properties of the hypergeometric functions we can rewrite the $d$-dimensional Legendre polynomials as
\begin{equation}
	P^{(d)}_\ell(x) = {\Gamma(1+\ell)\Gamma(d-3) \over \Gamma(\ell+d-3)} C^{({d-3 \over 2})}_\ell(x) = \frac{C^{({d-3 \over 2})}_\ell(x)}{C^{({d-3 \over 2})}_\ell(1)} ,
\end{equation}
where $C_\ell^{(d)}$ are the Gegenbauer polynomials. Notice that $P^{(d)}_\ell(1)=1$.
The coefficients $c_\ell^{(d)}$ in the decomposition \eqref{eq:decomposition_partial_amplitudes} read as
\begin{equation}
	\label{eq:coefs_c}
	c_\ell^{(d)} = \frac{2 \,\pi ^{\frac{d}{2}-1} (d+2 \ell-3) \Gamma (d+\ell-3)}{\Gamma \left(\frac{d}{2}-1\right) \Gamma (\ell+1)}\,,
\end{equation}
For further discussion of the above expressions see either \cite{Karateev:2019ymz} or \cite{Correia:2020xtr}. Notice that in equation \eqref{eq:factor_N} we have an additional factor of $2^{d-1}$ compared to the conventions of \cite{Correia:2020xtr}.

Unitarity of theory leads to the following non-linear constraint on the partial amplitudes 
\begin{equation}
	\label{eq:unitarity_NL}
	\forall s\geq 0,\quad
	\forall \ell:\qquad
	|\mathcal{S}_\ell(s)|^2\leq 1.
\end{equation}
In this paper we refer to the condition \eqref{eq:unitarity_NL} as the non-linear unitarity. This condition is equivalent to
\begin{equation}
	\label{eq:unitarity}
	2 \mathcal{N}_d(s)\,\text{Im}\mathcal{T}_\ell(s) \geq |\mathcal{T}_\ell(s)|^2.
\end{equation}
The unitarity constraint \eqref{eq:unitarity} can be written in a positive semi-definite form as 
\begin{equation}
	\label{eq:unitarity_practice}
	\begin{pmatrix}
		1&1\\
		1&1
	\end{pmatrix} + \frac{1}{\mathcal{N}_d(s)} 
	\begin{pmatrix}
		0& -i\mathcal{T}_\ell^*(s)	\\
		i\mathcal{T}_\ell(s)&0
	\end{pmatrix}\succeq 0
\end{equation}
or equivalently 
\begin{equation}
	\label{eq:unitarity_practice_2}
	\begin{pmatrix}
		1&0\\
		0&0
	\end{pmatrix} +
	\begin{pmatrix}
		-\frac{1}{2 \mathcal{N}_d}\Im \cT_\ell&\,\, \mathcal{N}_d^{-1/2}\Re\cT_\ell	\\[2pt]
		\mathcal{N}_d^{-1/2}\Re\cT_\ell&2\Im\cT_\ell
	\end{pmatrix}\succeq 0
\end{equation}
The condition \eqref{eq:unitarity_practice_2} has an advantage of consisting of purely real matrices compared to \eqref{eq:unitarity_practice}.

The non-linear unitarity condition \eqref{eq:unitarity} contains the following simple condition
\begin{equation}
	\label{eq:positivity}
\text{Im}\mathcal{T}_\ell(s) \geq 0.
\end{equation}
It is called positivity. The positivity condition is a necessary, but not sufficient condition to have a unitary theory.

\section{Effective amplitude from unitarity }
\label{app:one-loop_unitarity}
Consider the following amplitude
\begin{multline}
	\label{eq:tree_amplitude}
	\mathcal{T}^\text{tree}(s,t,u) = 
	g_2 \, (s^2+t^2+u^2) + g_3\, stu + g_4 \left(s^2+t^2+u^2\right)^2\\
	+ g_5\, s t u\left(s^2+t^2+u^2\right)
	+ g_6 \left(s^2+t^2+u^2\right)^3 + g_6' \left(s t u\right)^2 + O(s^7),
\end{multline}
where $g_n$ are real coefficients.
This amplitude arises in a typical tree level computation of the scattering of Goldstone bosons described by the Lagrangian density \eqref{eq:EFT_lagrangian}. The tree level amplitude \eqref{eq:tree_amplitude} does not satisfy \eqref{eq:unitarity}. One needs to compute loop corrections in order to obtain the full amplitude which obeys \eqref{eq:unitarity}. Let us write this full amplitude in the following form
\begin{equation}
	\label{eq:full_amplitude}
	\mathcal{T}^\text{full}(s,t,u) = 
	\mathcal{T}^\text{tree}(s,t,u) + N(s,t,u).
\end{equation}
Here the tree level amplitude in the small energy expansion is given by \eqref{eq:tree_amplitude} and the function $N(s,t,u)$ represents all the loop corrections.
Loop computation is a hard task. Luckily there is a simpler way to determine the function $N(s,t,u)$ as we show in this appendix.

The main idea is to use \eqref{eq:unitarity} on the full amplitude \eqref{eq:full_amplitude}, thus we get the following condition
\begin{equation}
	\label{eq:unitarity_full}
	\text{Im}N_\ell(s) = 2^{-d} s^\frac{d-4}{2} \big|\mathcal{T}^\text{full}_\ell(s)\big|^2 + \text{(particle production)}\,,
\end{equation}
where we have denote the partial wave projection of the function $N(s,t,u)$ by 
\begin{equation}\label{eq:projection_N}
	N_\ell(s) \equiv \int_{-1}^{+1} dx \mu_{d,\ell}(x) N(s,t(x),u(x)).
\end{equation}
We solve this equation iteratively order by order in small energy expansion by simply requiring non-linear unitarity \eqref{eq:unitarity}.\footnote{We remind that the small energy expansion means the expansion around $s=0$ where the variables $t$ and $u$ are written in the form \eqref{eq:tu_expressions} and the cosine of the scattering angle $x\equiv \cos\theta$ is kept fixed.} Off course, this procedure stops at the order in which particle production appears as emphasize in  \eqref{eq:unitarity_full}.

\paragraph{Leading order computation}
Let us start by focusing on the very first term in \eqref{eq:tree_amplitude}, namely we simply consider the following tree level amplitude
\begin{equation}
	\mathcal{T}^\text{tree}(s,t,u) = 
	g_2 \, (s^2+t^2+u^2) + O(s^3).
\end{equation}
According to \eqref{eq:PartialWaveDef} we can compute the partial amplitudes.
Only $\ell=0$ and $\ell=2$ components are non-zero. They read
\begin{equation}
	\label{eq:tree_projection}
		\mathcal{T}^\text{tree}_{\ell=0}(s)=a_0\, g_2 s^2\,\left(1+O(s)\right),\qquad
		\mathcal{T}^\text{tree}_{\ell=2}(s)=a_2\, g_2 s^2\,\left(1+O(s)\right),
\end{equation}
where the kinematic coefficients read as
\begin{equation}
	a_0\equiv\frac{3d-2}{ 2^{d}\pi^{(d-3)/2}\Gamma((d+1)/2)},\qquad
	a_2\equiv\frac{1}{ 2^{d}\pi^{(d-3)/2}\Gamma((d+3)/2)}.
\end{equation}
We can obtain the leading order of the right-hand side in \eqref{eq:unitarity_full} by \eqref{eq:tree_projection}, we get the following result then
\begin{equation}
	\label{eq:N_projection_tree}
		\text{Im}N^\text{tree}_{\ell=0}(s)=2^{-d}a_0^2\, g_2^2 s^{\frac{d}{2}+2}\,\left(1+O(s)\right),\qquad
		\text{Im}N^\text{tree}_{\ell=2}(s)=2^{-d}a_2^2\, g_2^2 s^{\frac{d}{2}+2}\,\left(1+O(s)\right).
\end{equation}

Analogously to \eqref{eq:decomposition_partial_amplitudes} we can write the function $N(s,t,u)$ as a sum of its partial waves, namely
\begin{equation}
	\label{eq:N_sum}
	N(s,t,u) = \sum_{\ell=0,2,\dots} c_\ell^{(d)} P_\ell^{(d)}\left(1+ \frac{2t}{s}\right)\,N_\ell(s).
\end{equation}
Plugging here the result \eqref{eq:N_projection_tree} and using \eqref{eq:definition_P} together with \eqref{eq:coefs_c} we finally get
\begin{equation}
	\label{eq:leading_imaginary}
	\text{Im}_sN(s,t,u)=s^{\tfrac{d}{2}} h(s|t,u),
\end{equation}
where the function $h(s|t,u)$ reads as
\begin{equation} h(s|t,u)\equiv\frac{g_2^2}{4^{d+1}\pi^{\tfrac{d-3}{2}}\Gamma(\tfrac{d+3}{2})}\Big((9d^2+6d-4)s^2+4(t^2+u^2)\Big)+O(s^3).
\end{equation}
Now we can simply guess the form of the function $N(s,t,u)$ which is crossing symmetric and satisfies the constraint \eqref{eq:leading_imaginary}. We get
\begin{equation}
	\label{eq:guess_N}
	N(s,t,u)=L_d(s)h(s|t,u)+L_d(t)h(t|s,u)+L_d(u)h(u|t,s),
\end{equation}
where the function $L_d(s)$ is defined as 
\begin{align}
	d=\text{even}:\qquad & L_d(s)\equiv-\frac{1}{\pi}\log(-s \sqrt{g_2})s^{\tfrac{d}{2}},\\
	d=\text{odd}:\qquad &  L_d(s)\equiv -(-s)^{\tfrac{d}{2}}.
\end{align}

\paragraph{The case of 4d}
Let us focus on $d=4$ for simplicity from now on. Let us summarize the result obtained in the previous section. The full amplitude in $d=4$ which obeys non-linear unitarity has the form \eqref{eq:full_amplitude}, where the function $N(s,t,u)$ has the following form 
\begin{equation}
	\label{eq:result_N_d=4}
N(s,t,u)=\log(-s \sqrt{g_2})f(s|t,u)+\log(-t \sqrt{g_2})f(t|s,u)+\log(-u \sqrt{g_2})f(u|t,s).
\end{equation}
The function $f(s|t,u)$ has the following small energy expansion
\begin{equation}
	\label{eq:expansion_f}
	 f(s|t,u)\equiv f_4(s|t,u)+f_5(s|t,u)+f_6(s|t,u)+O(s^6),\qquad f_n(s|t,u)\sim s^n,
\end{equation}
and the leading term in this expansion has just been found precisely and reads
\begin{equation}\label{eq:imf4}
	f_4(s|t,u)\equiv-\frac{g_2^2}{480\pi^2}s^2\Big(41\,s^2+t^2+u^2\Big).
\end{equation}
This is precisely the result  quoted in \eqref{eq:L_expression}.

\paragraph{The case of 6d}
As another example let us also write the effective amplitude in 6d. It reads
\begin{equation}
	\label{eq:amplitude_EFT_HD}
	\mathcal{T}_\text{EFT}(s,t,u) = 
	\mathcal{T}^{(2)}_\text{EFT}(s,t,u) + \mathcal{T}^{(3)}_\text{EFT}(s,t,u)+\mathcal{T}^{(4)}_\text{EFT}(s,t,u) +\mathcal{T}^{(5)}_\text{EFT}(s,t,u) + O(s^6),
\end{equation}
where we have
\begin{equation}
	\mathcal{T}^{(2)}_\text{EFT}= g_2 \, (s^2+t^2+u^2),\qquad
	\mathcal{T}^{(3)}_\text{EFT} = g_3\, stu,\qquad
	\mathcal{T}^{(4)}_\text{EFT} = g_4 \left(s^2+t^2+u^2\right)^2,
\end{equation}
together with 
\begin{multline}
	\label{eq:logs_d=6}
	\mathcal{T}^{(5)}_\text{EFT}(s,t,u) = g_5\, s t u\left(s^2+t^2+u^2\right)  - \frac{g_2^2}{26880\pi^3} \Big(
	s^3 (89 s^2+t^2+u^2) \log\left(-s\sqrt{g_2} \right) +\\
	t^3 (s^2+89 t^2+u^2) \log\left(-t\sqrt{g_2} \right) +
	u^3 (s^2+t^2+89 u^2) \log\left(-u\sqrt{g_2} \right) \Big).
\end{multline}

\paragraph{Computation of sub-leading orders in $d=4$}
We can compute the sub-leading terms in the expansion \eqref{eq:expansion_f} by recursively repeating the logic of the previous section. We do not provide further details and simply quote the final result. At the order $O(s^5)$ we get
\begin{equation}\label{eq:imf5}
f_5(s|t,u) =
-\frac{g_2g_3}{1920 \pi^2}s^3(36s^2-4(t^2+u^2)).
\end{equation}
At the order $O(s^6)$ we we get
\begin{equation}
	f_6(s|t,u) =  f_6^{(1)}(s|t,u)+ f_6^{(2)}(s|t,u)+ f_6^{(3)}(s|t,u),
\end{equation}
where the three functions are given by
\begin{align}
f_6^{(1)}(s|t,u) & = -\frac{g_3^2}{1920\pi^2}s^4(s^2+t^2+u^2),\\
f_6^{(2)}(s|t,u) & =-\frac{g_2 g_4}{560\pi^2}s^4(79 s^2+4(t^2+u^2)),
\end{align}
together with
\begin{multline}
	f_6^{(3)}(s|t,u)=\frac{g_2^3}{48384000 \pi^4}\times\\
	s^4\Big(7560\log(s)\big(79s^2+4(t^2+u^2)\big)+2561(t^2+u^2)-39029 s^2\Big).
\end{multline}
Where we notice that the later correspond to a two-loop computation.

\section{Weak coupling amplitudes and absence of bounds}
\label{app:absence_bounds}
Following \cite{Caron-Huot:2020cmc}, we can define the following two amplitudes in $d$ space-time dimensions
\begin{align}
	\label{eq:amplitude_1}
	\cT_{\text{spin 0}}(s,t,u)&\equiv-\lambda ^2 M^{4-d}
	\left(\frac{M^2}{s-M^2}+\frac{M^2}{t-M^2}+\frac{M^2}{u-M^2}+3\right),\\
	\label{eq:amplitude_2}
	\cT_{\text{stu pole}}(s,t,u)&\equiv-\lambda^2  M^{4-d}\left(\frac{M^6}{\left(s-M^2\right) \left(t-M^2\right)
		\left(u-M^2\right)}+1\right) - \gamma(d)\cT_{\text{spin 0}}(s,t,u),
\end{align}
where we have defined
\begin{equation}
	\label{eq:definition_gamma}
	\gamma(d) \equiv \frac{4}{9}\, {}_2F_1\left( \frac{1}{2}, 1, \frac{d-1}{2}; \frac{1}{9} \right)\,,
\end{equation}
to remove the spin $0$ contribution to $\cT_{\text{stu pole}}$.
Both amplitudes trivially satisfy crossing and maximal analyticity. 

Let us now expand these amplitudes around small energies. This can be done by using \eqref{eq:tu_expressions} and expanding around $s=0$. 
For the first amplitude we obtain
\begin{multline}
	M^d\, \mathcal{T}_{\text{spin 0}}(s,t,u) = M^{2}\lambda^2 (s+t+u)+\lambda^2(s^2+t^2+u^2)+M^{-2}\lambda^2 (s^3+t^3+u^3)+\\
	M^{-4}\lambda^2(s^4+t^4+u^4) + O(s^5).
\end{multline}
Comparing this expression with \eqref{eq:soft_1}, we conclude that for the first amplitude we have
\begin{equation}
	\label{eq:g_amplitude1}
	\cT_{\text{spin 0}}:\qquad
	g_2 = M^{-d} \lambda^2,\qquad
	g_3 = 3M^{-d-2} \lambda^2,\qquad
	g_4 = \frac{1}{2}\,M^{-d-4} \lambda^2.
\end{equation}
Analogously we can obtain the low energy coefficients for the second amplitude, they read
\begin{equation}
	\label{eq:g_amplitude2}
	\cT_{\text{stu pole}}:\qquad
	\begin{aligned}
	g_2 &= \frac{1}{2}M^{-d} \lambda^2\big(1-2\gamma(d)\big),\\
	g_3 &= M^{-d-2} \lambda^2\big(1-3\gamma(d)\big),\\
	g_4 &= \frac{1}{4}\,M^{-d-4} \lambda^2\big(1-2\gamma(d)\big).
	\end{aligned}
\end{equation}
Plugging the results \eqref{eq:g_amplitude1} and \eqref{eq:g_amplitude2} into the definition of dimensionless couplings  \eqref{eq:observables} we obtain the following expressions
\begin{align}
	\label{eq:gb_amplitude1}
	&\cT_{\text{spin 0}}:\qquad
	\bar g_3 = \frac{3}{\lambda^{4/d}},\qquad
	\bar g_4 = \frac{1}{2\lambda^{8/d}},\\
	\label{eq:gb_amplitude2}
	&\cT_{\text{stu pole}}:\qquad
	\bar g_3 = \frac{2^{1+\frac{2}{d}}(1-3\gamma(d))}{\lambda^{4/d}(1-2\gamma(d))^{1+\frac{2}{d}}},\qquad
	\bar g_4 = \frac{2^{\frac{4}{d}-2}}{\lambda^{8/d}(1-2\gamma(d))^{4/d}}.
\end{align}
We summarize the above result in a compact form in table  \ref{tab:coefficients_g}.
\begin{table}[h!]
	\centering
	\begin{tabular}{c|ccc|cc}
		\toprule
		& $g_2$&$g_3$&$g_4$&$\bar{g}_3$&$\bar{g}_4$\\
		\midrule
		$\cT_{\text{spin 0}}$&$\frac{\lambda^2}{M^d}$&$\frac{3\lambda^2}{M^{d+2}}$&$\frac{\lambda^2}{2M^{d+4}}$&$\frac{3}{\lambda^{4/d}}$&$\frac{1}{2\lambda^{8/d}}$\\[5pt]
		$	\cT_{\text{stu pole}}$&$\frac{\lambda^2(1-2\gamma(d))}{2M^d}$&$\frac{\lambda^2(1-3\gamma(d))}{M^{d+2}}$&$\frac{\lambda^2(1-2\gamma(d))}{4M^{d+4}}$&$\frac{2^{1+\frac{2}{d}}(1-3\gamma(d))}{\lambda^{4/d}(1-2\gamma(d))^{1+\frac{2}{d}}}$&$\frac{2^{\frac{4}{d}-2}}{\lambda^{8/d}(1-2\gamma(d))^{4/d}}$\\
		\bottomrule
	\end{tabular}
\caption{The low energy expansion coefficients for the amplitudes \eqref{eq:amplitude_1} and \eqref{eq:amplitude_2}.}
\label{tab:coefficients_g}
\end{table}

For generic values of the coupling $\lambda$ the amplitudes \eqref{eq:amplitude_1} and \eqref{eq:amplitude_2} do not satisfy the requirement of the non-linear unitarity. However, in the extreme weakly coupling limit $\lambda\rightarrow 0$ this problem is alleviated and unitarity is reduced to the positivity of the residues. In this sense, both amplitudes satisfied ``weak coupling'' unitarity \cite{Caron-Huot:2020cmc}.

Let us now discuss if any bounds on the parameter $\bar g_3$ can be constructed. From \eqref{eq:gb_amplitude1} we observe that the amplitude $\cT_{\text{spin 0}}$ allows for infinitely large values of $\bar g_3$ in any number of dimensions, namely
\begin{equation}
	\lim_{\lambda\rightarrow 0} \bar g_3 = 
	\lim_{\lambda\rightarrow 0} \frac{3}{\lambda^{4/d}} = 
	+\infty.
\end{equation}
This means that there always exists a weakly coupled theory with a very large positive value of $\bar g_3$, thus by construction no upper bound exists for this quantity. Analogously we can conclude that there is no lower bound on $\bar g_3$ in $d>2$ space-time dimensions. This follows from the $\bar g_3$ expression for the $\cT_{\text{stu pole}}$ amplitude in \eqref{eq:gb_amplitude2}, namely
\begin{equation}
	\lim_{\lambda\rightarrow 0} \bar g_3 = 
	\lim_{\lambda\rightarrow 0} \frac{2^{1+\frac{2}{d}}(1-3\gamma(d))}{\lambda^{4/d}(1-2\gamma(d))^{1+\frac{2}{d}}} = 
	-\infty.
\end{equation}
The latter equality holds in $d>2$ and can be straightforwardly checked by using the definition \eqref{eq:definition_gamma}.
Also notice, that from the definition of $\gamma(d)$ \eqref{eq:definition_gamma}, it follows that $g_2\geq 0$ for the amplitude $\cT_{\text{stu pole}}$ as required by positivity.

Rerunning these arguments for the $\bar g_4$ parameter we observe that no upper bound on $\bar g_4$ can be constructed. However, no statement can be made about the lower bound, which is consistent with our numerical result presented in figure \ref{fig:bound}.

It is interesting to discuss the absence of bounds on $\bar g_3$ in the context of the 6d $a$-theorem. In \cite{Elvang:2012st} the authors used the massless probe field (often ambiguously called the dilaton) to measure the difference between the UV and the IR $a$-anomaly. This difference is denoted by $\Delta a$. By using the Weyl anomaly matching they obtained the low energy behaviour of the scattering amplitudes of the probe field. Their result is given by equations (3.18) and (3.19)  in \cite{Elvang:2012st}. We can connect their notation with ours by comparing equations (3.18) and (3.19)  in \cite{Elvang:2012st} with \eqref{eq:soft_1}. We find that
\begin{equation}
	\label{eq:relating_notation}
	g_2 = \frac{\bar b}{f^6},\qquad
	g_3 = \frac{9\Delta a}{2f^8},\qquad
	\bar b \equiv \frac{b}{2f^2},
\end{equation}
where $\bar b$ is some dimensionless coefficient and $f>0$ is a dimensionful parameter. From positivity one concludes that $\bar b\geq 0.$
Plugging \eqref{eq:relating_notation} into \eqref{eq:observables} in $d=6$ we obtain
\begin{equation}
	\bar g_3 = \frac{9}{2}\,\bar b^{-4/3}\,\Delta a.
\end{equation}
Since we have just shown that by construction no bounds exist on $\bar g_3$ one concludes that it is impossible to prove the 6d $a$-theorem by simply considering the 2-to-2 scattering of the probe fields.

\newpage
\section{Complex spin partial amplitude}\label{sec:complexSpinSl}
\begin{figure}[h!]
	\centering
	\includegraphics[width=\linewidth]{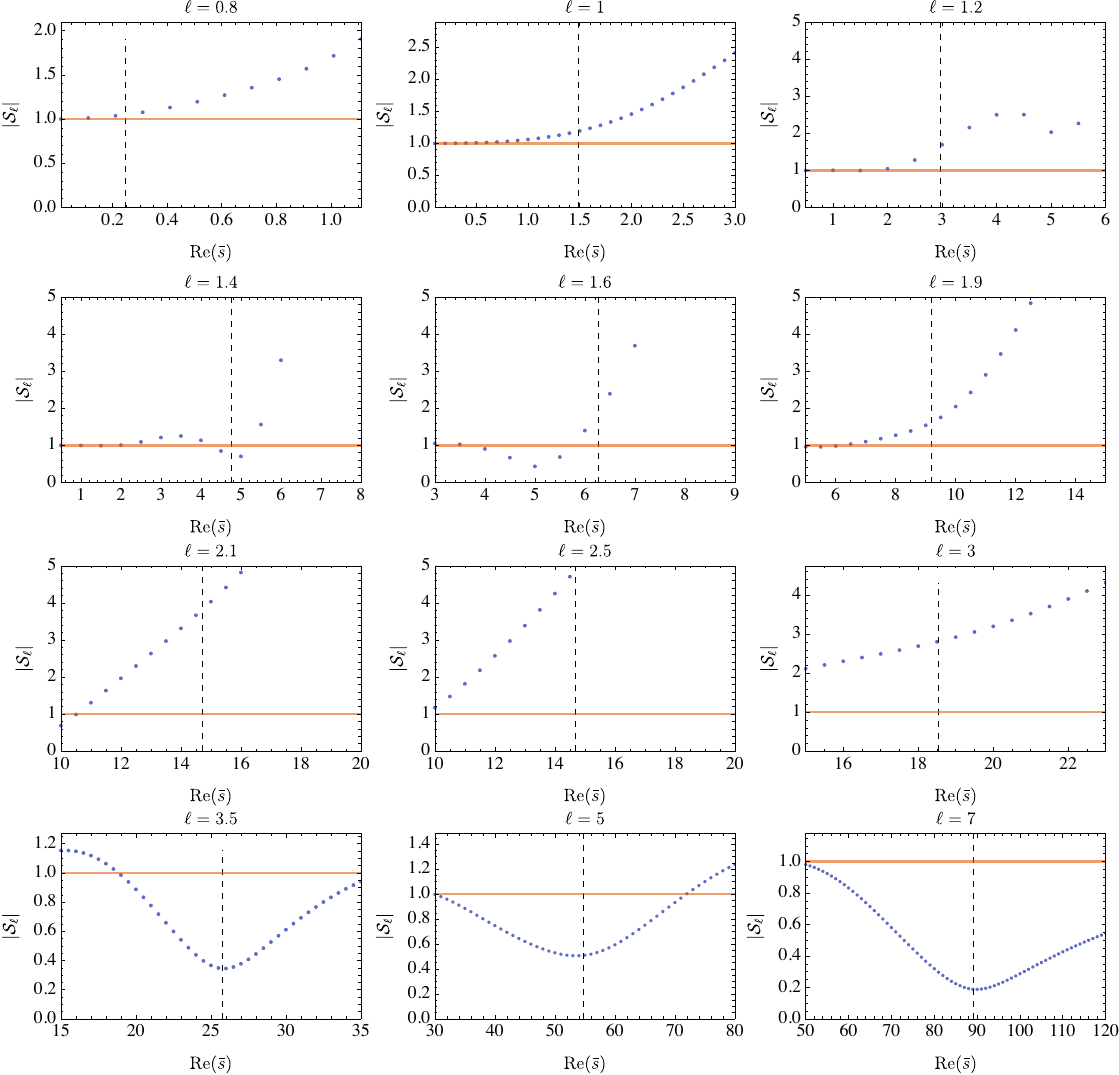}
	\caption{Absolute value of the partial amplitude $\mathcal{S}_\ell$ for complex spin on the real axis. The data are represented in blue. $|\mathcal{S}_\ell|=1$ is plotted in orange for reference. The position of the leading resonance of the leading trajectory is indicated by a black dashed line.  }
	\label{fig:complexSpinSl}
\end{figure}

\newpage
\section{Inverse amplitude method}
\label{app:IAP}

Using elastic unitarity it is possible to unitarize tree-level amplitudes in the $s$-channel. One method for doing this is called the Inverse Amplitude Method (IAM) (see for example \cite{Truong:1988zp,Dobado:1996ps}). The main idea is to \emph{assume} elastic unitarity which implies
\begin{equation}\label{eq:IAM_elastic}
	2\cN_d(s)\Im \cT_\ell(s) = |\cT_\ell(s)|^2 \qquad \Rightarrow \qquad \Im\left( \frac{1}{\cT_\ell(s)} \right) = -\frac{1}{2\cN_d(s)}.
\end{equation}
This equation fixes the imaginary part of the amplitude and leaves the real part free. At this point we are left to choose how the imaginary part arises. The simplest way to solve the above equation is by requiring
\begin{equation}
\frac{1}{\cT_\ell(s)} =  -\frac{i}{2\cN_d(s)} + \frac{1}{\alpha_\ell(s)}\,, \qquad \alpha_\ell(s)\in\mathbb{R}.
\end{equation}
Equivalently, this can be rewritten as
\begin{equation}
	\label{eq:IAM_ansatz}
	\text{IAM 1:}\qquad
	\mathcal{S}_\ell(s) = \frac{1+ i \frac{\alpha_\ell(s)}{2 \cN_d(s)}}{1- i \frac{\alpha_\ell(s)}{2 \cN_d(s)}}\,.
\end{equation}
The partial amplitude \eqref{eq:IAM_ansatz} is clearly a pure phase and, thus, obeys elastic unitarity.  The real function $\alpha_\ell(s)$ is called the \emph{seed}. Its explicit form depends on the model under consideration. We refer to the partial amplitude \eqref{eq:IAM_ansatz} as the amplitudes obtained with the IAM 1.

In the case of massless particles, we know that the imaginary part arises from the $\log(-s)$ term. Using this information we can, thus, make another choice for solving \eqref{eq:IAM_elastic}. Using $\Im (\log(-s)) = -\pi$ for $s>0$ one can rewrite \eqref{eq:IAM_elastic} as
\begin{equation}
	\Im\left( \frac{1}{\cT_\ell(s)} \right) = \Im\left( \frac
	{\log(-s)}{2\pi\cN_d(s)} \right).
\end{equation}
Solving this equation we obtain 
\begin{equation}
	\label{eq:IAM_ansatz2}
	\text{IAM 2:}\qquad
	\cT_\ell(s) = \frac{\tilde \alpha_\ell(s)}{1+ \tilde \alpha_\ell(s) \frac
		{\log(-s)}{2\pi\cN_d(s)}}\,,\qquad  \tilde \alpha_\ell(s)\in \mathbb{R}\,.
\end{equation}
We refer to the interacting part of the partial amplitude \eqref{eq:IAM_ansatz2} as the one obtained with the IAM 2.
The two solutions \eqref{eq:IAM_ansatz} and \eqref{eq:IAM_ansatz2} are equivalent upon the following redefinition of the seeds
\begin{equation}
	\frac{1}{\alpha_\ell(s)} = 	\frac{1}{\tilde \alpha_\ell(s)}  + \frac{\log(s)}{2\cN_d(s) \pi}\,.
\end{equation}

 The natural choice for the seeds reads as
\begin{align}
	\text{IAM 1} :\quad \alpha_\ell(s) = \cT_\ell^{tree}(s),\\
	\text{IAM 2} : \quad \tilde \alpha_\ell(s) = \cT_\ell^{tree}(s).
\end{align}
In figure  \ref{fig:spin0_IAMvsNum} we have plotted the spin zero partial amplitudes obtained using  IAM1 and IAM2 for the three-level amplitude \eqref{eq:tree_amplitude_main}. Both IAM1 and IAM2 expressions have a similar behaviour. 
The advantage of IAM 2 is that it naturally reproduces the correct low energy behaviour \eqref{eq:soft_1} up to $O(s^4)$ order.

\section{Sum-rules for the coefficients $g_n$}
\label{app:sum-rules}

The coefficients $g_n$ are defined in \eqref{eq:soft_1}. There are relation which relate the value of these coefficients to certain integral over the imaginary part of the interacting part of the scattering amplitude. We refer to this relations as the sum-rules.

The most efficient way to derive this sum-rules is by using the technology of \cite{Bellazzini:2020cot}. There the authors introduced the notion of arcs $a_n(s,t)$. They are defined as
\begin{equation}
	\label{eq:arcs}
	a_n(s,t) \equiv \frac{1}{2\pi i} \oint_{\mathcal{C}_s} \frac{ds'}{ s'} \frac{\cT(s',t)}{[s'(s'+t)]^{n+1}}\,,
\end{equation}
where $\mathcal{C}_s$ is the circle centered at $-t/2$ and of radius $s+t/2$ and exclude the real axis. For $n\geq 0$ we can deform the contour and drop the arc at infinity provided that the amplitude admit 2 subtractions.\footnote{In gapped theory, this is guarantied due to the Martin-Froissart bound \cite{Martin:1965jj,Jin:1964zza}.  See also \cite{Haring:2022cyf} for a recent discussion for massless particles. }
\begin{equation}\label{key}
	a_n(s,t) = \frac{1}{\pi} \int_s^\infty ds' \left(\frac{1}{s'} + \frac{1}{s'+t}\right) \frac{\Im T(s',t)}{[s'(s'+t)]^{n+1}}.
\end{equation}
The arcs \eqref{eq:arcs} are related to the $g_n$ coefficients. Indeed, by consider a contour $\mathcal C_s$ of small radius, we can use $\mathcal{T}_{\text{EFT}}$ in \eqref{eq:arcs}. For example, the first arc is given by
\begin{equation}\label{eq:arc_a0}
	a_0(s,t) = 2 g_2 - g_3 t  + O(s^2) 
\end{equation}
The two simplest sum-rules follows directly and reads
\begin{align}
	\label{eq:sum-rule_g2}
	g_2 &= \frac{1}{\pi}  \int_0^\infty ds \frac{\Im \cT(s,0)}{s^3},\\
	\label{eq:sum-rule_g3}
	g_3 &= \frac{2}{\pi}  \int_0^\infty ds \left(\frac{3\Im \cT(s,t)}{2s^4} -\frac{\partial_t\Im \cT(s,t)}{s^3}\right)_{t\to0}.
\end{align}
Using the positivity constraint \eqref{eq:positivity} one immediately concludes that $g_2\geq 0$. No similar statement can be made about $g_3$. The above sum-rule  for $g_3$ exists only if the first derivative in $t$ in the forward limit, namely $\partial_t\Im \cT(s,t=0)$, is finite. This not required by basic axioms. However, it is true in all example we encounter. It would be interesting to understand if this statement can be proven.  
From the perspective of the arc \eqref{eq:arc_a0}, this is equivalent to the statement that the $O(s^2)$ satisfy the same property. 

It is very useful to decompose the amplitudes in \eqref{eq:sum-rule_g2} and \eqref{eq:sum-rule_g3} into partial amplitudes according to \eqref{eq:decomposition_partial_amplitudes}. We can then define the following coefficients
\begin{align}
	\label{eq:sum-rule_g2_spin}
	g_2^{(\ell)} &\equiv \frac{c_\ell^{(d)}}{\pi}\int_0^\infty ds \frac{\Im\cT_\ell(s)}{s^3},\\
	\label{eq:sum-rule_g3_spin}
	g_3^{(\ell)} &\equiv \frac{2 c_\ell^{(d)}}{\pi}\int_0^\infty ds \frac{\Im\cT_\ell(s)}{s^4}\left(\frac{3}{2} - \frac{2 \ell (d+\ell-3)}{(d-2)}\right).
\end{align}
Since we work with the scattering of identical particles $\ell=0,2,4,\ldots$
Comparing these to \eqref{eq:sum-rule_g2} and \eqref{eq:sum-rule_g3} we conclude that
\begin{equation}
	g_n = \sum_{\ell=0}^\infty g_n^{(\ell)},\qquad n=2,3.
\end{equation}
In order to see this we used the following equalities
\begin{equation}
	P_\ell^{(d)}(1) = 1 \,, \qquad \partial_t P_\ell^{(d)}\left(1+ \frac{2t}{s}\right)\Bigg|_{t\to 0} = \frac{2 j (d+\ell-3)}{(d-2)} \frac{1}{s},
\end{equation}
which are straightforward to obtain from the explicit expression \eqref{eq:definition_P}.

For completeness let us also write the sum-rule for the coefficient $g_4$. It reads as
\begin{equation}
	g_4= \frac{1}{2\pi}\int_{0}^\infty ds \left(\frac{\Im T(s,0)}{s^5} + \frac{42 \pi \beta}{s(1+s \sqrt{g_2})}\right),
\end{equation}
where  $\beta = -\frac{g_2^2}{480\pi^2}$ in $d=4$ and $\beta=0$ in $d\geq 5$.
In order to derive this expression we compute the arc $a_1$ explicitly in the forward limit to obtain
\begin{equation}
	g_4 + 21 \beta \log\left(s \sqrt{g_2}\right) + O\left(s\right) = \frac{1}{2\pi}  \int_{s}^\infty ds \frac{\Im T(s,0)}{s^5}
\end{equation}
to simplify, we can make use of the identity
\begin{equation}
	\log(x) = \int_{x}^{\infty}dy\frac{-1}{y(1+y)} + \cO(x).
\end{equation}
To write 
\begin{align}
	g_4 &= \frac{1}{2\pi}\int_{s}^\infty ds \left(\frac{\Im T(s,0)}{s^5} + \frac{42 \pi \beta}{s(1+s \sqrt{g_2})}\right) + O\left(s \right)\\
	&= \frac{1}{2\pi}\int_{0}^\infty ds \left(\frac{\Im T(s,0)}{s^5} + \frac{42 \pi \beta}{s(1+s  \sqrt{g_2})}\right)
\end{align}

\section{Large energy constraints}
\label{app:large_energy}

In this appendix we study the large energy behaviour $s\rightarrow \infty$ of the simplest ansatz \eqref{eq:ansatz_1} reviewed in section \ref{sec:review}. Let us write this ansatz here again for convenience
\begin{equation}
	\label{eq:ansatz_HD}
	\mathcal{T}_\text{ansatz}(s,t,u) = \sum_{a,b,c}
	\alpha_{abc}\, \rho^a(s)\rho^b(t)\rho^c(u) +  \mathbb{N}(s,t,u).
\end{equation}
The term $\mathbb{N}(s,t,u)$ contains the logs. Its precise form depends on the dimension. In what follows we will completely ignore $\mathbb{N}(s,t,u)$ for simplicity. However, once $\mathbb{N}(s,t,u)$ is known the discussion below can be straightforwardly adopted in order to take it into account.  
The $\rho$-variables were defined in \eqref{eq:rho_definition}. 
In the $s\rightarrow \infty $ limit the $\rho$-variables have the following expansions
\begin{equation}
	\label{eq:expansion_rho}
	\begin{aligned}
		\rho(s) &= - 1 + \frac{2 i}{s^{1/2}} + \frac{2}{s} - \frac{2i}{s^{3/2}} + O\left(s^{-2}\right),\\
		\rho(t) &= - 1 + \frac{2\sqrt{2}}{(1-x)^{1/2}}\,\frac{1}{s^{1/2}}- \frac{4}{1-x}\,\frac{1}{s}+
		+ \frac{4\sqrt{2}}{(1-x)^{3/2}}\,\frac{1}{s^{3/2}} + O\left(s^{-2}\right),\\
		\rho(u) &= - 1 + \frac{2\sqrt{2}}{(1+x)^{1/2}}\,\frac{1}{s^{1/2}}- \frac{4}{1+x}\,\frac{1}{s}+
		+ \frac{4\sqrt{2}}{(1+x)^{3/2}}\,\frac{1}{s^{3/2}} + O\left(s^{-2}\right),
	\end{aligned}
\end{equation}
where $x$ is defined as the cosine of the scattering angle $\theta$, see \eqref{eq:observables}. Plugging these into the ansatz \eqref{eq:ansatz_HD} we can write the following expansions
\begin{equation}
	\label{eq:expansion_ansatz}
	\mathcal{T}_\text{ansatz}(s,t,u) = 
	A_0+\frac{A_1(x)}{s^{1/2}}+\frac{A_2(x)}{s}+\frac{A_3(x)}{s^{3/2}}+O(s^{-2}),
\end{equation}
where the coefficients $A_n$ are straightforward to compute. Let us write for concreteness the first two coefficient in this expansion, they read
\begin{equation}
	\label{eq:coefficients_A}
	\begin{aligned}
		A_0     &= \sum_{a,b,c}\alpha_{abc}\, (-1)^{a+b+c},\\
		A_1(x) &= \sum_{a,b,c}-2\,\alpha_{abc}\, (-1)^{a+b+c}\left(i a+\frac{\sqrt{2}b}{\sqrt{1-x}}+\frac{\sqrt{2}c}{\sqrt{1+x}}\right).
	\end{aligned}
\end{equation}
The coefficient $A_0$ is real instead the coefficients $A_n(x)$ with $n\geq 1$ have both real and imaginary parts.

The expansion \eqref{eq:expansion_ansatz} can then be used to evaluate the large energy behaviour of the partial amplitude related to the scattering amplitude via \eqref{eq:PartialWaveDef}. Let us write this expression here for convenience
\begin{equation}
	\label{eq:PA_HD}
	\mathcal{T}_\ell(s) = \frac{\ell!\,\Gamma\left(\frac{d-3}{2}\right)}{4\,\pi^{(d-1)/2}\Gamma(d-3+\ell)}
	\times \int_{-1}^{+1} dx \big((1-x)(1+x)\big)^{\frac{d-4}{2}} C_\ell^{(d-3)/2}(x) \mathcal{T}(s,t(x),u(x)),
\end{equation}
where $C_\ell^{(d-3)/2}(x)$ is the Gegenbauer polynomial. Plugging \eqref{eq:expansion_ansatz} into \eqref{eq:PA_HD} we obtain
\begin{equation}
	\label{eq:expansion_PA_HD}
	\mathcal{T}_\ell(s) = B^\ell_0+\frac{B^\ell_1}{s^{1/2}}+\frac{B^\ell_2}{s}+\frac{B^\ell_3}{s^{3/2}}+O(s^{-2}),
\end{equation}
where the coefficients $B^\ell_n$ are real numbers given by
\begin{equation}
	\label{eq:coefficient_Bjn}
	B^\ell_n \equiv \frac{\ell!\,\Gamma\left(\frac{d-3}{2}\right)}{4\,\pi^{(d-1)/2}\Gamma(d-3+\ell)}
	\times \int_{-1}^{+1} dx \big((1-x)(1+x)\big)^{\frac{d-4}{2}} C_\ell^{(d-3)/2}(x) A_n(x).
\end{equation}
Let us define for convenience the following constants
\begin{equation}
	m \equiv \frac{2^{2-d}\pi^\frac{3-d}{2}}{\Gamma\left(\frac{d-1}{2}\right)},\qquad
	n_\ell\equiv\frac{2^{d-7/2}\Gamma\left(\ell+1/2\right)\Gamma\left(\frac{d-1}{2}\right)\Gamma\left(\frac{d-3}{2}\right)}{\pi \Gamma\left(d+\ell-5/2\right)}.
\end{equation}
Then plugging \eqref{eq:coefficients_A} into \eqref{eq:coefficient_Bjn} we get
\begin{equation}
	\label{eq:values_Bj}
	\begin{aligned}
		B^\ell_0  &= m\delta_{\ell,0} \sum_{a,b,c}\alpha_{abc}\, (-1)^{a+b+c},\\
		B^\ell_1(x) &= m\sum_{a,b,c}-2\,\alpha_{abc}\, (-1)^{a+b+c}\left(i a \delta_{\ell,0}+\sqrt{2}(b+c)n_\ell\right).
	\end{aligned}
\end{equation}

Notice that one should worry about commuting the expansion \eqref{eq:expansion_ansatz} with the integration in \eqref{eq:PA_HD}. The practical way of testing if the expression \eqref{eq:expansion_PA_HD} is correct is to evaluate the coefficients $B^\ell_n$ defined in \eqref{eq:coefficient_Bjn}. We quickly observe that $B^\ell_n$ remain finite only for 
\begin{equation}
	\label{eq:limit_convergence}
	n\leq d/2.
\end{equation} 
As a result the representation \eqref{eq:expansion_PA_HD} breaks down for $n>d/2$. This happens because of the presence of $x=\pm 1$ singularities in \eqref{eq:expansion_rho} which are not compensated in the integrals \eqref{eq:coefficient_Bjn}.

Now recall the non-linear unitarity condition on partial amplitudes given by \eqref{eq:unitarity} together with \eqref{eq:factor_N}. Let us write it here explicitly for convenience
\begin{equation}
	\label{eq:unitarity_higher_dimensions}
	\text{Im}\mathcal{T}_\ell(s) \geq 2^{-d} s^{\frac{d-4}{2}}\, |\mathcal{T}_\ell(s)|^2.
\end{equation}
Plugging here the expansion \eqref{eq:expansion_PA_HD} and taking into account that the coefficient $B^\ell_0$ is real as a consequence of the reality of the coefficient $A_0$, we obtain
\begin{equation}
	\label{eq:condition_large_energy}
	\frac{\text{Im}B^\ell_1}{s^{1/2}}+\frac{\text{Im}B^\ell_2}{s}+\frac{\text{Im}B^\ell_3}{s^{3/2}}+O(s^{-2}) \geq 
	2^{-d} s^{\frac{d-4}{2}}\, \left| B^\ell_0+\frac{B^\ell_1}{s^{1/2}}+\frac{B^\ell_2}{s}+\frac{B^\ell_3}{s^{3/2}}+O(s^{-2})\right|^2.
\end{equation}
In order for this inequality to be satisfied we need to require that the left-hand side decays with $s$ at the same rate or slower than the right-hand side. This imposes the constraints on the coefficients $B_n^\ell$. Below we derive these constraints for several values of $d$.

\paragraph{The case of 4d.} In the case of $d=4$ the condition \eqref{eq:condition_large_energy} reads as
\begin{equation}
	\begin{aligned}
		\ell=0:\qquad \frac{\text{Im}B^0_1}{s^{1/2}}+\frac{\text{Im}B^0_2}{s}+O(s^{-3/2}) &\geq 
		2^{-d} \, \left| B^0_0+\frac{B^0_1}{s^{1/2}}+\frac{B^0_2}{s}+O(s^{-3/2})\right|^2,\\
		\ell\geq2:\qquad \frac{\text{Im}B^\ell_1}{s^{1/2}}+\frac{\text{Im}B^\ell_2}{s}+O(s^{-3/2}) &\geq 
		2^{-d} \, \left|\frac{B^\ell_1}{s^{1/2}}+\frac{B^\ell_2}{s}+O(s^{-3/2})\right|^2.
	\end{aligned}
\end{equation}
Recall that $B^\ell_0=0$ for $\ell\geq 2$ due to \eqref{eq:values_Bj}. The above conditions can be satisfied only if we require $B_0^0=0$. Taking into account \eqref{eq:values_Bj} we conclude that in 4d the following condition must be obeyed
\begin{equation}
	d=4:\qquad
	\sum_{a,b,c}\alpha_{abc}\, (-1)^{a+b+c} = 0.
\end{equation}
Requiring this condition is equivalent to setting $A_0=0$ in the $s\rightarrow \infty$ expansion of the ansatz \eqref{eq:expansion_ansatz}.

\paragraph{The case of 5d.} In the case of $d=5$ the condition \eqref{eq:condition_large_energy} reads as
\begin{equation}
	\begin{aligned}
		\ell=0:\qquad \frac{\text{Im}B^0_1}{s^{1/2}}+\frac{\text{Im}B^0_2}{s}+O(s^{-3/2}) &\geq 
		2^{-d}s^{1/2} \, \left| B^0_0+\frac{B^0_1}{s^{1/2}}+\frac{B^0_2}{s}+O(s^{-3/2})\right|^2,\\
		\ell\geq2:\qquad \frac{\text{Im}B^\ell_1}{s^{1/2}}+\frac{\text{Im}B^\ell_2}{s}+O(s^{-3/2}) &\geq 
		2^{-d}s^{1/2} \, \left|\frac{B^\ell_1}{s^{1/2}}+\frac{B^\ell_2}{s}+O(s^{-3/2})\right|^2.
	\end{aligned}
\end{equation}
The above conditions can be again satisfied only if we require $B_0^0=0$. Taking into account \eqref{eq:values_Bj} we conclude that in 5d the same condition as in $d=4$ must be obeyed, namely
\begin{equation}
	d=5:\qquad
	\sum_{a,b,c}\alpha_{abc}\, (-1)^{a+b+c} = 0.
\end{equation}

\paragraph{The case of 6d.} In the case of $d=6$ the condition \eqref{eq:condition_large_energy} reads as
\begin{equation}
	\begin{aligned}
		\ell=0:\qquad \frac{\text{Im}B^0_1}{s^{1/2}}+\frac{\text{Im}B^0_2}{s}+O(s^{-3/2}) &\geq 
		2^{-d}s \, \left| B^0_0+\frac{B^0_1}{s^{1/2}}+\frac{B^0_2}{s}+O(s^{-3/2})\right|^2,\\
		\ell\geq2:\qquad \frac{\text{Im}B^\ell_1}{s^{1/2}}+\frac{\text{Im}B^\ell_2}{s}+O(s^{-3/2}) &\geq 
		2^{-d}s \, \left|\frac{B^\ell_1}{s^{1/2}}+\frac{B^\ell_2}{s}+O(s^{-3/2})\right|^2.
	\end{aligned}
\end{equation}
The above conditions can be satisfied only if we require $B_0^0=0$ and $B_1^\ell=0$ for any $\ell\geq 0$. Taking into account \eqref{eq:values_Bj} we conclude that in 6d the following conditions must be obeyed
\begin{equation}
	\label{eq:conditions_6d}
	d=6:\qquad
\begin{aligned}
	&\sum_{a,b,c}\alpha_{abc}\, (-1)^{a+b+c} = 0,\\
	&\sum_{a,b,c}\alpha_{abc}\, a(-1)^{a+b+c} = 0,\\
	&\sum_{a,b,c}\alpha_{abc}\, (b+c)(-1)^{a+b+c} = 0.
\end{aligned}
\end{equation}
Notice, that since $\alpha_{abc}$ are totally symmetric, the last condition in \eqref{eq:conditions_6d} is actually redundant.
Requiring these conditions is equivalent to setting $A_0=0$ and $A_1(x)=0$ in the $s\rightarrow \infty$ expansion of the ansatz \eqref{eq:expansion_ansatz}.

\bibliographystyle{JHEP}
\bibliography{refs}

\end{document}